\documentclass[%
 reprint,
superscriptaddress,
 amsmath,amssymb,
 aps,
prb,
floatfix,
]{revtex4-2}

\usepackage{xcolor}
\usepackage{graphicx}
\graphicspath{{/Users/pit.bermes/Documents/t-J model/plots/paper_magnons}}
\usepackage{dcolumn}
\usepackage{bm}
\usepackage{hyperref}

\begin{document}
\title{Magnetic polarons beyond linear spin-wave theory: Mesons dressed by magnons}

\author{Pit Bermes}
\affiliation{%
 Department of Physics and Arnold Sommerfeld Center for Theoretical Physics (ASC),
Ludwig-Maximilians-Universit\"at M\"unchen, Theresienstr. 37, M\"unchen D-80333, Germany
}%
\affiliation{%
 Munich Center for Quantum Science and Technology (MCQST), Schellingstr. 4, D-80799 M\"unchen, Germany
}%

\author{Annabelle Bohrdt}
\affiliation{University of Regensburg, Universit\"atsstr. 31, Regensburg D-93053, Germany}
\affiliation{%
 Munich Center for Quantum Science and Technology (MCQST), Schellingstr. 4, D-80799 M\"unchen, Germany
}%

\author{Fabian Grusdt}%
\affiliation{%
 Department of Physics and Arnold Sommerfeld Center for Theoretical Physics (ASC),
Ludwig-Maximilians-Universit\"at M\"unchen, Theresienstr. 37, M\"unchen D-80333, Germany
}%
\affiliation{%
 Munich Center for Quantum Science and Technology (MCQST), Schellingstr. 4, D-80799 M\"unchen, Germany
}%

\date{\today}

\begin{abstract}
When a mobile hole is doped into an antiferromagnet, its movement will distort the surrounding magnetic order and yield a magnetic polaron. The resulting complex interplay of spin and charge degrees of freedom gives rise to very rich physics and is widely believed to be at the heart of high-temperature superconductivity in cuprates.
In this paper, we develop a quantitative theoretical formalism, based on the phenomenological parton description, to describe magnetic polarons in the strong coupling regime. We construct an effective Hamiltonian with weak coupling to the spin-wave excitations in the background, making the use of standard polaronic methods possible.
Our starting point is a single hole doped into an AFM described by a ‘geometric string’ capturing the strongly correlated hopping processes of charge and spin degrees of freedom, beyond linear spin-wave approximation. Subsequently, we introduce magnon excitations through a generalized 1/S expansion and derive an effective coupling of these spin-waves to the hole plus the string (the meson) to arrive at an effective polaron Hamiltonian with density-density type interactions. After making a Born-Oppenheimer-type approximation, this system is solved using the self-consistent Born approximation to extract the renormalized polaron properties. We apply our formalism (i) to calculate beyond linear spin-wave ARPES spectra, (ii) to reveal the interplay  of ro-vibrational meson excitations, and (ii) to analyze the pseudogap expected at low doping. Moreover, our work paves the way for exploring magnetic polarons out-of equilibrium or in frustrated systems, where weak-coupling approaches are desirable and going beyond linear spin-wave theory becomes necessary.
\end{abstract}

\maketitle


\section{Introduction}\label{sec:introduction}

Ever since the discovery of superconductivity in cuprates by Bednorz et al.~\cite{Bednorz1986} in 1986, tremendous effort has been put into understanding the rich physics of these materials. Today, there is general agreement that the physics underlying the cuprates is that of a doped antiferromagnetic (AFM) Mott insulator (MI). The Hubbard model as well as its strong coupling limit, the $t$-$J$ model, have emerged as minimal models capturing the strongly correlated processes of these materials \cite{Lee2006}. They feature many properties and different states of matter which have remained elusive for more than forty years in spite of the numerous studies on doped MI's.

A necessary step in order to reach a deeper understanding of the underlying mechanism of these phases is to understand and characterize the charge carriers in these systems. At very low doping, the charge carriers are quasiparticles named magnetic polarons and are formed by holes dressed by spin fluctuations~\cite{Sachdev1989, Kane1989}, a view recently corroborated by high-resolution ARPES measurements in clean cuprates~\cite{Kurokawa2023} and in ultracold atom experiments~\cite{Koepsell2019}. The magnetic frustration surrounding the hole is due to the competition of the antiferromagnetic order and the delocalization of the holes, favouring ferromagnetic order, and has been proposed to provide a possible pairing mechanism for high-temperature superconductivity~\cite{Emery1987,Schrieffer1988}. Therefore understanding and characterizing the magnetic polaron has been of great interest over the last few decades, but even the problem of a single polaron can be quite challenging, especially in the strong coupling regime $t > J$ relevant for high-temperature cuprate superconductors for which typically $t/J \approx 3$~\cite{Lee2006}.

At strong coupling, perturbative methods break down and different methods have been applied so far to this problem with varying success. Due to the complexity of the problem, large-scale numerical methods such as exact diagonalization \cite{Dagotto1990, Wang2021, Bonca1989, Hasegawa1989}, various Monte Carlo methods \cite{Brunner2000, Mishchenko2001, Blomquist2020, Boninsegni1991} and Density Matrix Renormalization Group (DMRG) methods \cite{White2001,Zhu2014,Wang2021, Bohrdt2020, Grusdt2019} have been applied, but most of them are restricted by system size or cannot access dynamical or out-of-equilibrium properties. Furthermore there have been variational approaches \cite{Sachdev1989}, and most prominently semi-analytical approaches using $1/S$-expansion and self-consistent Born approximation (SCBA) \cite{Kane1989, Liu1992, Schmitt-Rink1988, Martinez1991, Nielsen2021, Igarashi1992}. In addition, different groups have analyzed string-like excitations caused by the hole movement \cite{Grusdt2018, Trugman1988, Brinkman1970, Manousakis2007, Bulaeski1968} and Refs. \cite{Grusdt2018, Grusdt2019} have used this geometric string picture to develop a microscopic parton model~\cite{Beran1996} to describe the magnetic polarons.
Moreover, there has recently again been increased interest in this field due to experimental progress in ultracold quantum gases and quantum gas microscopes, leading to the ability to experimentally realize the Fermi-Hubbard model \cite{Cheuk2015, Parsons2016, Boll2016, Bohrdt2021coldatoms}. These experiments give new insights and possibilities to test theoretical models. They have already been able to measure the microscopic structure of magnetic polarons \cite{Koepsell2019}, signatures of string patterns \cite{Chiu2019, Bohrdt2019} and dynamical polaron formation \cite{Ji2021} using snapshots with single-site and single-atom resolution.

Despite multiple decades of studies, several aspects of magnetic polarons remain poorly understood. This concerns mostly their excited states and, in extension, their out-of- equilibrium properties, e.g. in transport measurements. Addressing these challenges is difficult within the established framework, and more concerningly it is unclear whether the conventional linear spin-wave model — the workhorse when it comes to describing the polaron ground state — remains accurate at higher energies. Concretely, the shape of the polaron spectrum at high energies remains debated and its shape at very low energies has not been analyzed in detail; the microscopic origin of the low-doping pseudogap between the nodal and antinodal point remains poorly understood; and the fate of magnetic polarons and their ro-vibrational excitations~\cite{Bohrdt2021} away from the dispersion minimum remains unclear. Finally, predicting far-from equilibrium dynamics and finite temperature properties of magnetic polarons~\cite{Hahn2022} represents an even greater challenge.

\begin{figure*}[t!]
  \centering\includegraphics[width=0.9\linewidth]{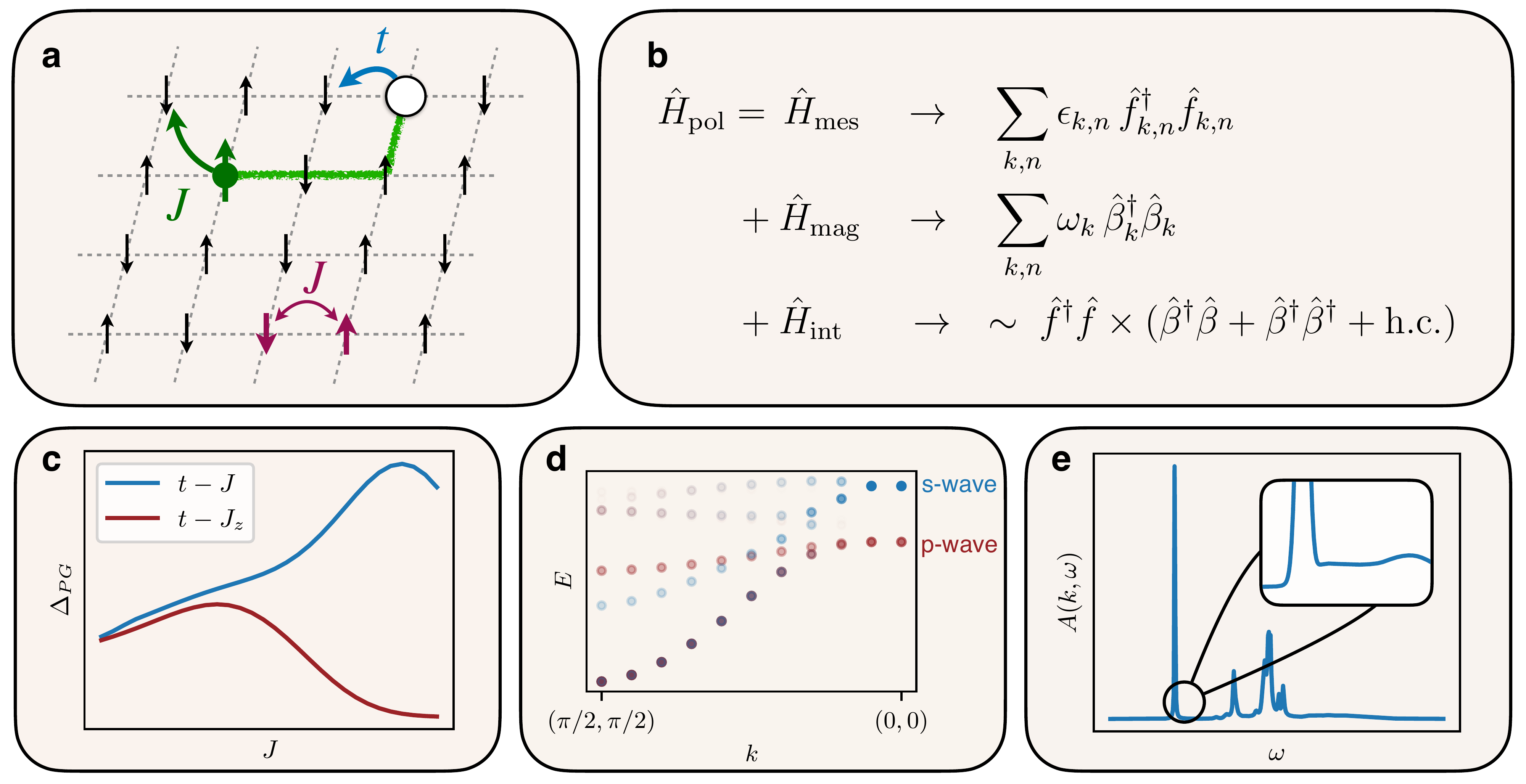}
  \caption{Overview of the main results. (a) We sketch the geometric string model for magnetic polarons. The hopping of the hole induces a string of flipped spins, and spin flip-flop processes lead to a dispersion of the meson as well as to the creation of spin-wave excitations away from the hole. In (b) we summarize the effective polaron Hamiltonian derived in the main text, describing meson-magnon coupling. An overview of our main results is shown in (c)-(e), starting with the pseudogap as a function of the spin coupling $J/t$ in (c). Panel (d) shows the crossing of different internal rotational excitations of the meson by choosing the opacity of the data points to be proportional to the overlap of the meson with rotational trial state \eqref{eq:rot-trial-states} and we plot the spectral function in (e).}
  \label{fig:overview}
\end{figure*}

In this article, we study the problem of a single magnetic polaron in a doped AFM-MI and construct an effective polaron Hamiltonian beyond linear spin-wave approximation. In this formalism, we use geometric strings to describe the strongly correlated hopping processes of the spin and charge degrees of freedom in a more complete and accurate way than the conventional 1/S expansion is able to achieve. Then, to improve the model, we also allow for collective spin-wave excitations and derive a coupling between the spin-waves and the bare polaron in the string picture. We thus obtain an effective Hamiltonian  for the magnetic polaron in the strong coupling regime as a central result of this work, see Fig.~\ref{fig:overview}(a),~(b). This effective Hamiltonian has only weak interactions, allowing for the use of many methods developed in the field of Bose polarons \cite{Devreese2015, Scazza2022, Grusdt2015} to obtain new insights in these systems. For example, we use the self-consistent Born approximation to solve the interacting Hamiltonian and benchmark our results against previous semi-analytical methods and against advanced numerical simulations.

The main advantages of our approach are two-fold: on one hand, we introduce a systematic scheme for describing magnetic polarons beyond the linear spin-wave approximation. The latter is known to overestimate coherent features in the single-hole spectral function \cite{Wrzosek2021} — so-called shake-off processes leading to a ladder-like spectrum of multiple resonances which were found to be absent beyond the first vibrational excitation in more accurate numerical studies of the t-J model \cite{Bohrdt2020, Brunner2000, Mishchenko2001}. On the other hand, we provide a quantitative description of the meson-picture of magnetic polarons \cite{Beran1996, Grusdt2018}, which loosely views them as confined mesonic pairs of a spinon and a chargon connected by a string and moving in a host AFM. We explicitly describe the meson part, going beyond linear spin-wave theory, and include a coupling of the meson to collective magnon excitations. I.e. the bare meson takes the role of a free impurity, which is weakly dressed with magnons, mostly via conventional density-density interactions. This allows us to give an estimate of the accuracy of the mesonic string model and check whether it captures all essential processes or gets strongly renormalized by the magnon dressing.

In this paper we employ our method to reveal several aspects of magnetic polarons that remained elusive so far, see Fig.~\ref{fig:overview}~(c)~-~(e). After benchmarking our approach by comparison of ground state energies to established Monte-Carlo results \cite{Brunner2000, Mishchenko2001}, we calculate the size of the anti-nodal (pseudo)gap at low-doping i.e. in the regime where magnetic polaron theory applies, and elucidate its microscopic origins. We further analyze the bare meson state and find indications for strong hybridization of the vibrational ground- with rotationally excited string states around the $\Gamma$ point at zero momentum, which is supported by DMRG simulations. Finally we calculate the magnetic polaron spectrum in which we reveal deviations from linear string theory \cite{Grusdt2018}; we also resolve the incoherent magnon contribution at very low excitation energies relevant to ARPES experiments in solids and confirm a peak-dip-hump feature below the super-exchange energy scale.

This article is organized as follows. In Sec. \ref{sec:formalsim} we construct the formalism from first principles. Starting from the microscopic $t$-$J$ model in Sec. \ref{subsec:model}, we use geometric strings in Sec. \ref{subsec:string_picture} to describe the most important and strongly correlated processes leading to string formation. In Sec. \ref{subsec:gen_expansion} we then include collective spin-wave excitations via the generalized 1/S-expansion and derive an effective coupling between the spin-waves and the string excitations in Sec. \ref{subsec:strong_coupling_approx}. The resulting effective Hamiltonian is then solved in the self-consistent Born approximation (SCBA) in Sec. \ref{subsec:SCBA}. Finally, in Sec. \ref{sec:results} we present the results obtained with this new method and discuss the polaron's renormalized properties, with particular focus on the (pseudo)gap at the anti-nodal point.

\section{Formalism}\label{sec:formalsim}
In this section, we will construct from first principle an effective model for magnetic polarons in an AFM valid at strong coupling. To this end, we will describe the magnetic polaron in the string-picture in section \ref{subsec:string_picture} as introduced in \cite{Grusdt2018, Trugman1988, Brinkman1970, Manousakis2007, Bulaeski1968}. In section \ref{subsec:gen_expansion}, we will use the generalized 1/S expansion \cite{Grusdt2018, Cubela2023} to include collective spin-wave excitations and compute the effective interaction between these spin-waves and the magnetic polaron. As a first application of the beyond-linear spin-wave Hamiltonian that we derive (summarized in Sec.~\ref{subsec:strategy}) we solve it using the self-consistent Born approximation (SCBA) in section \ref{subsec:SCBA}.

\subsection{Model}\label{subsec:model}
We start from the $t-J$ model \cite{Auerbach1998} in the 2D square lattice
\begin{align}
\label{t-j-Hamiltonian}
\hat{H}=&-t\sum_{\langle i,j\rangle}\sum_{\sigma}\hat{P}\left(\hat{c}^{\dagger}_{i\sigma}\hat{c}_{j\sigma}+h.c.\right)\hat{P}\nonumber\\
&+J\sum_{\langle i,j\rangle}\left(\hat{\mathbf{S}}_i \cdot \hat{\mathbf{S}}_j-\frac{1}{4}\hat{n}_{i}\hat{n}_{j}\right) ,
\end{align}
where $\hat{P}$ denotes the projector onto the subspace with no more than one fermion $\hat{c}_{j\sigma}$ per lattice site. $\hat{\mathbf{S}}_j$ and $\hat{n}_{j}$ denote the spin and density operators at site $j$, $\sum_{\langle i,j\rangle}$ is the sum over all nearest-neighbor bonds $\langle i,j \rangle$ where each bond is counted once and $\sigma=\uparrow,\downarrow$.
In the following, we will only consider the one-hole subspace, so that the last term in \eqref{t-j-Hamiltonian} is constant and does not affect the ground state.

Next, we use a Schwinger boson representation of Eq.~\eqref{t-j-Hamiltonian} by introducing a spinless fermionic chargon operator $\hat{h}^{\dagger}_{j}$ and Schwinger bosons $\hat{b}^{\dagger}_{j\sigma}$, such that we can write the electron operator as $\hat{c}^{\dagger}_{j\sigma}=\hat{h}_{j}\hat{b}^{\dagger}_{j\sigma}$ for the spin-$1/2$ case \footnote{Note that the statistics of $\hat{h}$ (bosonic or fermionic) in the considered case with just one hole on a bipartite lattice does not affect the physics. Our choice of fermionic $\hat{h}$ and Schwinger bosons $\hat{b}$ instead of bosonic $\hat{h}$ and fermionic $\hat{b}$ is convenient but not necessary for what follows.}, see e.g. \cite{Auerbach1998}.
For general spin $S$ of the fermion $\hat{c}_{j\sigma}$, the single occupancy constraint due to the projector $\hat{P}$ can be generalized to \cite{Grusdt2018, Cubela2023}

\begin{equation}
\label{constraint}
\sum_{\sigma}\hat{b}^{\dagger}_{j\sigma}\hat{b}_{j\sigma}=2S\left(1-\hat{h}^{\dagger}_{j}\hat{h}_{j}\right).
\end{equation}
This constraint ensures that if there is a hole at a lattice site $j$, then there will be no spin at this site. Note that for $S\neq \frac{1}{2}$ this constraint is different from the conventional constraint $\sum_{\sigma}\hat{b}^{\dagger}_{j\sigma}\hat{b}_{j\sigma}+\hat{h}^{\dagger}_{j}\hat{h}_{j}=2S$ usually used in $1/S$ expansions \cite{Kane1989}, but the former constraint allows to describe non-linear distortions of the local Néel order parameter.
Using these representations (for $S=\frac{1}{2}$) \footnote{for general $S$, the chargon hopping term becomes $-t\sum_{\langle i,j\rangle }\sum_{\sigma}\left(\hat{\mathcal{F}}_{ij}(S)\hat{h}_{i}\hat{h}^{\dagger}_{j}+h.c.\right)$ where $\hat{\mathcal{F}}_{ij}$ describes the reordering of the spins during the hopping process and is necessary to fulfill the constraint \eqref{constraint}} and generalizing to anisotropic spin coupling, i.e. $J\,\hat{\mathbf{S}}_i \cdot \hat{\mathbf{S}}_j \rightarrow J_{\perp}(\hat{S}_i^x \hat{S}_j^x + \hat{S}_i^y \hat{S}_j^y) + J_{z}\hat{S}_i^z\hat{S}_j^z$, the Hamiltonian \eqref{t-j-Hamiltonian} becomes
\begin{align}
\label{Hamiltonian_schwinger}
\hat{H}=&-t\sum_{\langle i,j\rangle }\sum_{\sigma}\left(\hat{h}_{i}\hat{b}^{\dagger}_{i\sigma}\hat{b}_{j\sigma}\hat{h}^{\dagger}_{j}+h.c.\right)\nonumber\\
&+\dfrac{J_{z}}{4}\sum_{\langle i,j\rangle }\left(\hat{b}^{\dagger}_{i\uparrow}\hat{b}_{i\uparrow}-\hat{b}^{\dagger}_{i\downarrow}\hat{b}_{i\downarrow}\right)\left(\hat{b}^{\dagger}_{j\uparrow}\hat{b}_{j\uparrow}-\hat{b}^{\dagger}_{j\downarrow}\hat{b}_{j\downarrow}\right)\nonumber\\
&+\dfrac{J_{\perp}}{2}\sum_{\langle i,j\rangle }\left(\hat{b}^{\dagger}_{i\uparrow}\hat{b}_{i\downarrow}\hat{b}^{\dagger}_{j\downarrow}\hat{b}_{j\uparrow}+\hat{b}^{\dagger}_{i\downarrow}\hat{b}_{i\uparrow}\hat{b}^{\dagger}_{j\uparrow}\hat{b}_{j\downarrow}\right)\\
=&\,\hat{H}_{t}+\hat{H}_{J_z}+\hat{H}_{J_\perp} .\nonumber
\end{align}

Using the translational invariance of \eqref{Hamiltonian_schwinger}, i.e. the conservation of the total momentum, we can follow Lee, Low and Pines \cite{Lee1953} and shift to the frame co-moving with the hole by applying the unitary transformation
\begin{equation}
\label{LLP}
\hat{U}_{LLP}=e^{-i\hat{\mathbf{X}}_{h}\cdot\hat{\mathbf{Q}}_{b}},
\end{equation}
where we introduced the chargon position operator ${\hat{\mathbf{X}}_{h}=\sum_{j}\mathbf{j}\,\hat{h}^{\dagger}_{j}\hat{h}_{j}}$ as well as the total Schwinger boson momentum operator $\hat{\mathbf{Q}}_{b}=\sum_{k}\sum_{\sigma}\mathbf{k}\,\hat{b}^{\dagger}_{k\sigma}\hat{b}_{k\sigma}$.

Applying this transformation as well as a Fourier transformation, the Hamiltonian becomes block-diagonal in the hole-momentum basis.
\begin{equation}
\hat{H} = \sum_{k}\hat{H}(k)\hat{h}_{k}^{\dagger}\hat{h}_{k}\, ,
\end{equation}
with
\begin{align}
\label{Hamiltonian_LLP}
\hat{H}(k)=&\langle k|\hat{U}_{LLP}^{\dagger}\;\hat{H}\;\hat{U}_{LLP}|k\rangle\nonumber\\
=\,&t\sum_{\delta\,\sigma}\left(e^{i(\hat{\mathbf{Q}}_{b}-k)\delta}\,\hat{b}^{\dagger}_{0,\sigma}\,\hat{b}_{\delta,\sigma}+h.c.\right)\,\nonumber\\
&+\dfrac{J_{z}}{4}\sum_{\langle i,j\rangle }\left(\hat{b}^{\dagger}_{i\uparrow}\hat{b}_{i\uparrow}-\hat{b}^{\dagger}_{i\downarrow}\hat{b}_{i\downarrow}\right)\left(\hat{b}^{\dagger}_{j\uparrow}\hat{b}_{j\uparrow}-\hat{b}^{\dagger}_{j\downarrow}\hat{b}_{j\downarrow}\right)\nonumber\\
&+\dfrac{J_{\perp}}{2}\sum_{\langle i,j\rangle }\left(\hat{b}^{\dagger}_{i\uparrow}\hat{b}_{i\downarrow}\hat{b}^{\dagger}_{j\downarrow}\hat{b}_{j\uparrow}+\hat{b}^{\dagger}_{i\downarrow}\hat{b}_{i\uparrow}\hat{b}^{\dagger}_{j\uparrow}\hat{b}_{j\downarrow}\right).
\end{align}
Here we introduced the chargon momentum eigenstate $|k\rangle = \hat{h}^\dagger_{k}|0\rangle$ and the sum over the primitive vectors ${\delta \in \lbrace e_{x}, e_{y}\rbrace}$.
In addition, we used that $\hat{h}_{i}\,\hat{\mathbf{X}}_{h}=\hat{h}_{i}\,\mathbf{i}\,$ since we are only considering the subspace of a single hole.

\subsection{Strategy \& main result:\\ Generalized 1/S approach \& polaron Hamiltonian}\label{subsec:strategy}
In the following, we will be interested in the strong coupling regime $t > J$ where the chargon motion takes place on a short time scale $\propto 1/t$ compared to the time scale of the spin fluctuations  $\propto 1/J$. Thus, we will first solve for the fast chargon motion in a Néel-ordered Ising background described by the $t-J_{z}$ model using a truncated string basis. Within this basis, we will add a subset of spin flip-flop processes $J_{\perp} \hat{S}^{+}_{i}\hat{S}^{-}_{j}$ to get a model containing the most dominant spin processes while still being solvable with numerical diagonalization.

Then, in order to describe the remaining spin processes, we will add magnon processes on top, where magnons are Holstein-Primakoff bosons $\hat{a}^{\dagger}_{j}$ defined with respect to the distorted spin configuration $\hat{\tau}_{j}^{z}$ as described  by the geometric string created by the hole motion,
\begin{equation}
\hat{S}_{j}^{z}=\hat{\tau}_{j}^{z}\left(S-\hat{a}^{\dagger}_{j}\hat{a}_{j}\right)\,.
\end{equation}
Using the separation of time scales to decouple the meson dynamics and spin fluctuations, as well as including the magnons in linear spin-wave theory, will yield the beyond-linear spin-wave polaron Hamiltonian:
\begin{align}\label{eq:pol-Hamiltonian}
\hat{H} &= \sum_{n, k}E_n(k)\hat{f}^\dagger_{n,k}\hat{f}_{n,k}\nonumber\\
&+ \sum_{p}\omega_{p}^{\textrm{mag}}\hat{\beta}_{p}^{\dagger}\hat{\beta}_{p}\nonumber\\
&+\sum_{n,n', k,p,q}B^{(1)}\;\hat{f}^\dagger_{n',k}\hat{f}_{n,k+p-q}\hat{\beta}_{p}^{\dagger}\hat{\beta}_{q}\nonumber\\
&+\sum_{n,n', k,p,q}\left(B^{(2)}\;\hat{f}^\dagger_{n',k}\hat{f}_{n,k+p+q}\hat{\beta}_{p}^{\dagger}\hat{\beta}_{q}^{\dagger} + \textrm{h.c.}\right)\,.
\end{align}
Here the coupling parameters $A,\,B$, depend on all momenta $k,p,q$ and internal meson states $n,\,n'$; explicit expressions can be found in App. \ref{app:coefficients}.
This Hamiltonian consists of a free meson ($\hat{f}_{n,k}$), a free spin-wave ($\hat{\beta}_{p}$) and an interaction part.
Note that the free meson Hamiltonian already contains all correlated hopping processes of the hole and spins, including all the interaction terms between the hole and spins present in the conventional 1/S expansion \cite{Kane1989, Liu1992, Schmitt-Rink1988, Martinez1991, Nielsen2021, Igarashi1992}. The interaction terms in Eq.~\eqref{eq:pol-Hamiltonian} therefore describe interactions between the hole and spin background beyond linear spin-wave theory.
The effective model for the magnetic polaron \eqref{eq:pol-Hamiltonian} is the main result of this work and forms the starting point for different new ways of treating the magnetic polaron by using methods developed in polaron physics.

\subsection{Magnetic polaron in the string-picture}\label{subsec:string_picture}
\begin{figure}[t]
  \centering\includegraphics[width=.9\linewidth]{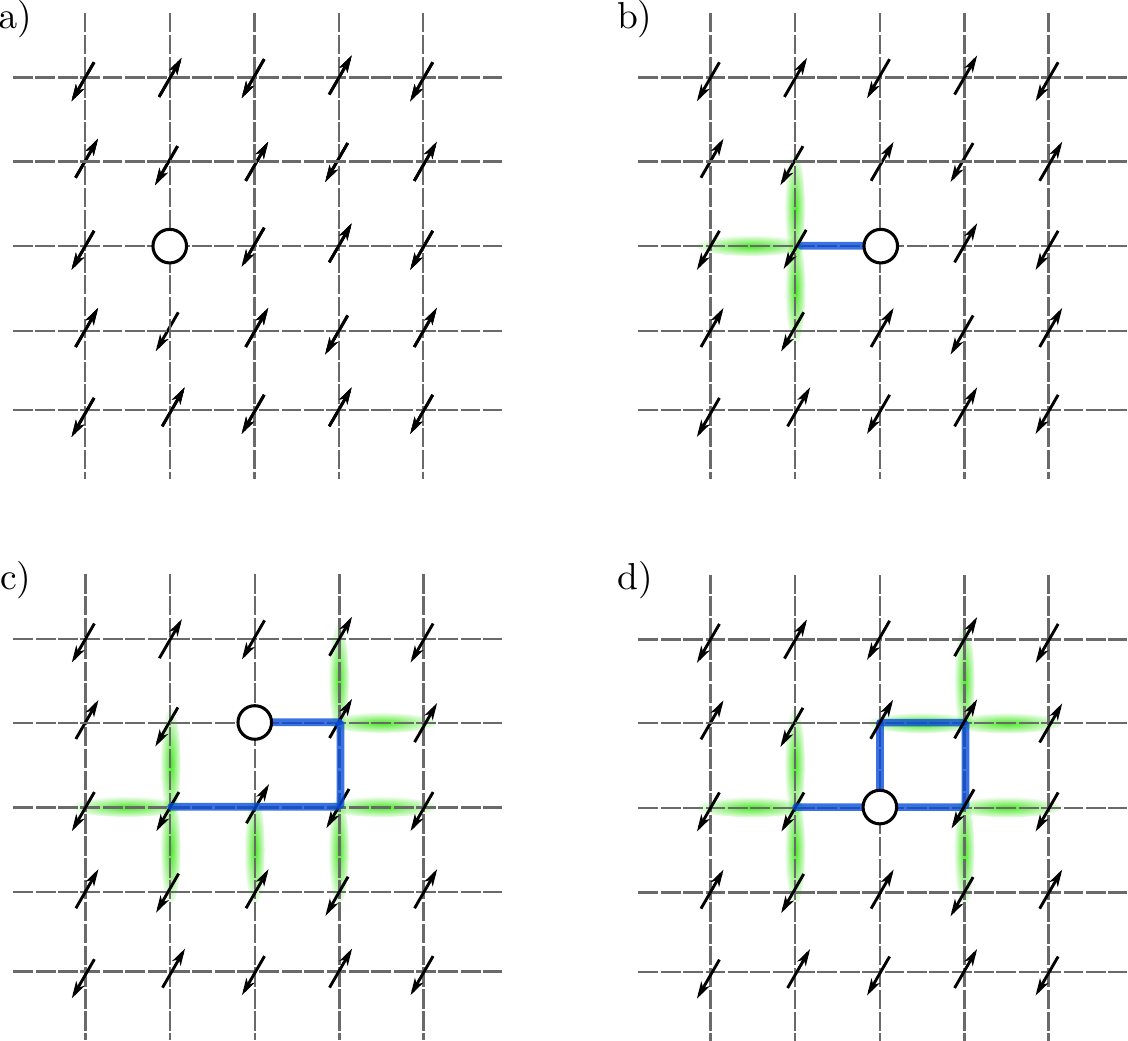}
  \caption{Illustration of the hole hopping in an AFM spin background and the induced magnetic frustration after 0(a), 1(b), 4(c) and 5(d) hopping processes. Panel (d) shows a state where the flipped spins do not form a connected graph and is therefore not included in the truncated basis. The path of the hole is colored blue while the frustrated bonds are highlighted in green.}
  \label{fig:hole_hopping}
\end{figure}

Now we derive the free meson part in Eq.~\eqref{eq:pol-Hamiltonian}.
Our starting point is a perfect Néel background with a single hole with momentum $k$ in the LLP frame. In the lab frame, when the hole hops to a neighboring site, the spin on this site hops in the opposite direction without changing its orientation and thus distorts the Néel order. In the strong coupling limit $t > J$, the hole can hop several times before the spins have time to adjust, which creates a "string" of distorted spins w.r.t. the Néel background as shown in Fig.~\ref{fig:hole_hopping}. In the LLP-frame, the hopping process produces the same effect with an additional translation applied afterwards shifting the hole back to the origin. As shown in \cite{Grusdt2018}, these strings can be viewed as binding two partons (spinon and chargon) and therefore we will in the following call the quasiparticles formed by the hole and its geometric string \emph{mesons} in analogy to \cite{Grusdt2018, Beran1996} and  high-energy physics.

Since longer strings have more frustrated bonds and therefore a higher energy cost, they are suppressed at low energies. This justifies the next approximation, where we will work in a truncated basis $\mathcal{B}_{trunc}=\lbrace | \alpha;\sigma,k\rangle\rbrace_{\alpha}$ consisting of physically distinct states with a string of length up to a certain $l_{max}$. $k$ and $\sigma$ label the LLP-momentum and the total spin $S^{z}_{tot}=-\sigma$, while $\alpha$ labels the string configuration. We construct this basis by starting from the classical Néel state with a hole doped into the origin $|0;\sigma,k\rangle =\hat{b}_{0\sigma}\hat{h}^{\dagger}_{k}|\textrm{Néel}\rangle$ and then applying the hopping term $\hat{H}_{t}(k)\big|_{\langle 0j\rangle }$ along sets of bonds consisting of up to $l_{max}$ segments. Here it is important to note that different strings can lead to the same spin configuration and should only be counted once in the truncated basis \cite{Trugman1988}. For instance, if the chargon travels along a Trugman loop, the shortest of them being going around a unit square one and a half times, the spin background is left undisturbed. More generally, every two trajectories differing only by a Trugman loop are equivalent and will give the same state, which we only count once.

This procedure creates a truncated basis of string states, and we emphasize that we do not apply $\hat{H}_{J_{\perp}}$ to add new states to the basis, in contrast to earlier approaches \cite{Manousakis2007}.
Furthermore, the goal of this string picture is to separate the spin dynamics due to the hole movement (energy scale $t$) from the dynamics due to the spin fluctuations (energy scale $J$). In order to better achieve this separation, we further truncate the basis and remove all string states where flipped spins and the hole do not form a connected graph; see Fig.~\ref{fig:hole_hopping}(d) for an example of such a string state. These states appear in our construction starting from $l_{max}=5$ because of strings intersecting themselves. But such states can be treated as consisting of a short string describing one of the connected components adjacent to the hole together with magnon excitations responsible for the remaining components which we will include in a subsequent step. For a discussion on this choice and the role of unconnected strings, see App.~\ref{app:connected-strings}.

Within the truncated basis, we can compute all matrix elements of \eqref{Hamiltonian_LLP} and  numerically diagonalize the resulting sparse matrix
\begin{equation}
\label{hamiltonitan_ED}
H^{ED}_{\alpha\beta}(k)=\langle \alpha|\hat{H}(k)|\beta\rangle ,
\end{equation}
where we omitted the spin and momentum labels $\sigma, k$ of the basis states for better readability.
Note that in truncating the physical Hilbert space, we are only considering transverse spin fluctuations $\propto J_{\perp}$ which connect one string configuration in the truncated basis to another, for instance by shortening the string. Other processes, which could for instance break up the string or would include bonds not adjacent to the string, are not included. Such processes will be added in the next step using linear spin-wave theory.

What we have achieved so far is a description of magnetic polarons in a truncated string basis. As in the case of a single hole in the $t-J_{z}$ model \cite{Grusdt2018}, this object can be viewed as a mesonic bound state of a confined spinon and a chargon. The main difference to the $t-J_{z}$ polaron is that we have included $J_{\perp}$-processes between the mesonic basis states, equipping it with additional spinon dynamics beyond Trugman loops \cite{Trugman1988}.

In preparation for the next step, where couplings to magnon excitations of the host quantum-Heisenberg AFM are included, we discuss how the total system momentum $\hat{Q}_b$ in the spin sector is distributed. The meson sector included in the truncated spin basis contributes $\hat{Q}_{\textrm{mes}}$; spin-waves added in the following section add an additional term $\hat{Q}_a$. I.e. the total momentum operator becomes
\begin{equation}\label{eq:momentum-decomp}
\hat{Q}_{b} = \hat{Q}_{\textrm{mes}} + \hat{Q}_{a}.
\end{equation}

In the truncated basis, the total magnon momentum $\hat{Q}_{a}$ can be treated as c-number.
Plugging the decomposition Eq.~\eqref{eq:momentum-decomp} into Eq.~\eqref{Hamiltonian_LLP} and using the meson eigenenergies $E_n(k)$ as well as the eigenstates $|\phi_n^{mes}(k)\rangle$ of \eqref{hamiltonitan_ED}, we can rewrite the Hamiltonian \eqref{Hamiltonian_LLP} as $\hat{H}(k) = \hat{H}_{mes}(k) + \hat{H}_{1}(k)$, where
\begin{equation}
\label{free1_polaron_hamiltonian}
\hat{H}_{mes}(k) = \sum_{n}E_{n}(k-\hat{Q}_{a})\, \hat{f}^{\dagger}_{n, k-\hat{Q}_{a}}\hat{f}_{n, k-\hat{Q}_{a}}\; .
\end{equation}
Here we defined the fermionic creation operator $\hat{f}^{\dagger}_{n, k}$ so that it creates one magnetic polaron with momentum $k$ in the $n$-th excited meson state: $\hat{f}^{\dagger}_{n, k}|0\rangle = |\phi_n^{mes}(k)\rangle$. The first part $\hat{H}_{mes}(k)$ contains all terms $\propto J_{z}$ acting on string states and the subset of terms $\propto J_{\perp}$ which can be described by geometric strings. The second part $\hat{H}_{1}(k)$ contains all remaining terms and is constructed in Sec.~\ref{subsec:gen_expansion}.

\subsection{Generalized 1/S expansion}\label{subsec:gen_expansion}
In this section we are going to use the generalized 1/S expansion as proposed in \cite{Grusdt2018} to include transverse spin fluctuations that are omitted in \eqref{hamiltonitan_ED}. Those terms $\hat{H}_{1, J_{\perp}}$ lead to the creation of spin-flip excitations (magnons), for which we now construct an effective magnon Hamiltonian. It contains two parts: a free spin-wave part and a coupling to the mesonic hole.

As usual we want to describe the quantum fluctuations using linear spin-wave theory corresponding to the Holstein-Primakoff (HP) approximation. But in order to include non-linear distortions of the classical Néel state from the motion of the chargon, we perform the HP approximation around this distorted state instead of the unperturbed Néel state.
The distortion created by the hole (or chargon) can be described by dynamical Ising variables $\hat{\tau}_{j}^{z}$ with
\begin{equation}
\label{ising variables}
\hat{\tau}_{j}^{z}=
\begin{cases}
+1\quad & \textrm{for} \quad |S\rangle_{j}\\
-1\quad & \textrm{for} \quad |-S\rangle_{j}
\end{cases}.
\end{equation}
Here $|\pm S\rangle_{j}$ denotes the two possible states a spin at site $j$ can have in the truncated basis. For $S=1/2$, the Ising variable  $\hat{\tau}_{j}^{z} = 2 \hat{S}_{j}^{z}$ is proportional to the local magnetization, see also Sec. IIB in \cite{Grusdt2018} for further details.
Furthermore, we set $\hat{\tau}_{X_{h}}^{z}=0$ at the position $\hat{X}_{h}$ of the chargon ($X_{h}=0$ in the LLP frame).

With the distorted background and the HP approximation, the spin operators are represented by:
\begin{align}
\label{eq:HP_operators}
\hat{S}_{j}^{z}&=\hat{\tau}_{j}^{z}\left(S-\hat{a}^{\dagger}_{j}\hat{a}_{j}\right),\nonumber\\
\hat{S}^{+\hat{\tau}_{j}^{z}}_{j}&=\sqrt{2S}\;\hat{a}_{j}\, ,\nonumber\\
\hat{S}^{-\hat{\tau}_{j}^{z}}_{j}&=\sqrt{2S}\;\hat{a}_{j}^{\dagger}\, .
\end{align}
To lowest order in $1/S$, the bosonic HP operators $\hat{a}_{j}$ are related to the Schwinger bosons by
\begin{align}
\hat{b}_{j,-\hat{\tau}_{j}^{z}}&=\hat{a}_{j}\, ,\nonumber\\
\hat{b}_{j,+\hat{\tau}_{j}^{z}}&=\sqrt{2S}\, .
\end{align}
Plugging this into \eqref{Hamiltonian_LLP}, we obtain
\begin{align}
\hat{H}_{J_{z}}&=J_{z}\sum_{0\notin\langle ij\rangle }\hat{S}^{z}_i \, \hat{S}^{z}_j\nonumber\\
&=J_{z}S^{2}\sum_{\langle ij\rangle }\hat{\tau}_{i}^{z}\hat{\tau}_{j}^{z} \underbrace{-J_{z}S\sum_{\langle ij\rangle }\hat{\tau}_{i}^{z}\hat{\tau}_{j}^{z}\left(\hat{a}_{i}^{\dagger}\hat{a}_{i}+\hat{a}_{j}^{\dagger}\hat{a}_{j}\right)}_{\equiv \hat{H}_{1, J_z}}\nonumber\\
&+\mathcal{O}(S^{0})\, .
\end{align}
The first term above, without magnon operators $\hat{a}_{j}$, is already included in the Hamiltonian \eqref{hamiltonitan_ED} which gets exactly diagonalized, so that we have to drop this term to avoid double-counting. Furthermore
\begin{align}
\label{eq:H1_jperp}
\hat{H}_{J_{\perp}}&=\dfrac{J_{\perp}}{2}\sum_{\langle ij\rangle }\left(\hat{S}_{i}^{+}\hat{S}_{j}^{-}+\hat{S}_{i}^{-}\hat{S}_{j}^{+}\right)\nonumber\\
&=\dfrac{1}{2}J_{\perp}S\sum_{0\notin\langle ij\rangle }\big[(1+\hat{\tau}_{i}^{z}\hat{\tau}_{j}^{z})(\hat{a}_{i}^{\dagger}\hat{a}_{j}+\hat{a}_{j}^{\dagger}\hat{a}_{i})\nonumber\\
&\qquad\qquad\qquad\; +(1-\hat{\tau}_{i}^{z}\hat{\tau}_{j}^{z})(\hat{a}_{i}\hat{a}_{j}+\hat{a}_{i}^{\dagger}\hat{a}_{j}^{\dagger}) \big] ,\nonumber\\
\hat{H}_{1,J_{\perp}}&=\dfrac{1}{2}J_{\perp}S\sum_{0\notin\langle ij\rangle }\big[(1+\hat{\tau}_{i}^{z}\hat{\tau}_{j}^{z})(\hat{a}_{i}^{\dagger}\hat{a}_{j}+\hat{a}_{j}^{\dagger}\hat{a}_{i})\nonumber\\
&\qquad +(1-\hat{\tau}_{i}^{z}\hat{\tau}_{j}^{z}-2\hat{F}_{ij})(\hat{a}_{i}\hat{a}_{j}+\hat{a}_{i}^{\dagger}\hat{a}_{j}^{\dagger}) \big].
\end{align}
Again, we have to avoid double counting processes already included on the level of the truncated basis. In the last line of \eqref{eq:H1_jperp}, we therefore subtract all contributions already accounted for in \eqref{hamiltonitan_ED} by defining the diagonal operator $\hat{F}_{ij}$:
\begin{align}
&\hat{F}_{ij}|\alpha\rangle\nonumber\\
&=
\begin{cases}
|\alpha\rangle\qquad\textrm{if }i,j\neq 0\;\text{and}\;|\alpha\rangle\in\mathcal{B}_{trunc}\\
\qquad\quad\;\text{and}\;(\hat{S}_{i}^{+}\hat{S}_{j}^{-}+\textrm{h.c.})|\alpha\rangle\in\mathcal{B}_{trunc}\\
0\qquad\quad\text{otherwise}
\end{cases}
\end{align}

Now, we would still have to consider the action of the hopping term $\hat{H}_{t}$ on the magnons. 
When the hole hops from a site $j$ onto a site $j+\delta$ occupied by a magnon, the magnon hops at the same time from $j+\delta$ to $j$. 
However, we neglect this kinetic interaction, since at our level of approximation, this will have only small effects on the long wavelength magnons, which dominate the dressing effects at low energies. A more detailed discussion of this can be found in Sec. \ref{subsec:strong_coupling_approx}.\\

In the following it will be useful to split the effective magnon Hamiltonian $\hat{H}_{1,J_{z}}+\hat{H}_{1,J_{\perp}}$ into the well known free magnon Hamiltonian $\hat{H}^{(0)}_{mag}$ describing free spin-waves in an undoped AFM and an interacting part $\hat{H}_{int}$ describing the influences of the hole and its disturbances of the background on the magnons:
\begin{align}
\hat{H}^{(0)}_{mag}&=J_{z}S\sum_{\langle ij\rangle }\left(\hat{a}_{i}^{\dagger}\hat{a}_{i}+\hat{a}_{j}^{\dagger}\hat{a}_{j}\right)\nonumber\\
&+J_{\perp}S\sum_{\langle ij\rangle }\left(\hat{a}_{i}\hat{a}_{j}+\hat{a}_{i}^{\dagger}\hat{a}_{j}^{\dagger}\right),\nonumber\\
\hat{H}_{int}&=-J_{z}S\sum_{\langle ij\rangle }\left(1+\hat{\tau}_{i}^{z}\hat{\tau}_{j}^{z}\right)\left(\hat{a}_{i}^{\dagger}\hat{a}_{i}+\hat{a}_{j}^{\dagger}\hat{a}_{j}\right)\nonumber\\
&+\dfrac{1}{2}J_{\perp}S\sum_{0\notin\langle ij\rangle }(1+\hat{\tau}_{i}^{z}\hat{\tau}_{j}^{z})\left(\hat{a}_{i}^{\dagger}\hat{a}_{j}+\hat{a}_{j}^{\dagger}\hat{a}_{i}\right)\nonumber\\
&-\dfrac{1}{2}J_{\perp}S\sum_{\langle ij\rangle }\left(1+\hat{\tau}_{i}^{z}\hat{\tau}_{j}^{z}+\delta_{i,0}+\delta_{j,0}+2\hat{F}_{ij}\right)\nonumber\\
&\qquad\qquad\qquad\times\left(\hat{a}_{i}\hat{a}_{j}+\hat{a}_{i}^{\dagger}\hat{a}_{j}^{\dagger}\right).
\end{align}
Note that far away from the hole, the spin background will not be disturbed and remain Néel ordered. Thus $\hat{\tau}_{i}^{z}\hat{\tau}_{j}^{z}=-1$ for nearest neighbors and the interaction will be localized around the hole. 

The free spin-wave Hamiltonian can be easily diagonalized using a Fourier transformation $\hat{a}_{j} = \frac{1}{\sqrt{N}}\sum_{k}e^{ikj}\hat{a}_{k}$ followed by a Bogoliubov transformation
\begin{equation}
\hat{a}_{p}=u_{p}\hat{\beta}_{p}-v_{p}\hat{\beta}_{-p}^{\dagger}\, ,
\end{equation}
where $\hat{\beta}_{p}$ are again bosonic operators. The coefficients $u_{p}, v_{p}$ are chosen such that this transformation diagonalizes $\hat{H}^{(0)}_{mag}$, so that we obtain
\begin{equation}
\label{free magnon hamiltonian}
\hat{H}^{(0)}_{mag} = \sum_{p}w_{p}^{mag}\hat{\beta}_{p}^{\dagger}\hat{\beta}_{p}+\dfrac{1}{2}\sum_{p}(w_{p}^{mag}-J_{z}Sz),
\end{equation}
with the well-known expressions \cite{Auerbach1998}
\begin{align}
w_{p}^{mag}&=zJ_zS\sqrt{1-\left(\frac{J_{\perp}}{J_{z}}\gamma_{p}\right)^2},\\
\gamma_{p}&=\frac{1}{z}\sum_{\delta}e^{ip\delta}\quad\text{where}\quad \delta \in \lbrace\pm e_{x},\pm e_{y}\rbrace,\\
u_{p}&=\sqrt{\frac{1}{2}\left(\frac{1}{\sqrt{1-(\frac{J_{\perp}}{J_{z}}\gamma_{p})^2}}+1\right)},\\
v_{p}&=\sqrt{\frac{1}{2}\left(\frac{1}{\sqrt{1-(\frac{J_{\perp}}{J_{z}}\gamma_{p})^2}}-1\right)}\text{sign}(\gamma_{p}).
\end{align}
Here, $z=4$ is the coordination number of the square lattice and the second sum in Eq.~\eqref{free magnon hamiltonian} simply gives the constant contribution from the spin fluctuations to the ground state energy of the undoped quantum Heisenberg AFM.

Note that the Bogoliubov transformation is ill-defined at the momenta $k = (0,0)$ and $k=(\pi,\pi)$ in the $SU(2)$-symmetric case where $J_{z} = J_{\perp}$. In the thermodynamic limit, all physical observables are well defined because the coefficients $u_p, v_p$ diverge as $|p|^{-1/2}$ for $p$ close to $(0,0)$ or $(\pi,\pi)$ and all integrals of quadratic expressions in these operators will give finite contributions.
For finite systems though, the HP approximation is invalid since the ground state does not break the $SU(2)$ symmetry and taking the zero modes into account gives unphysical divergent expressions. Therefore, in our numerical implementations where we have to approximate the integral as finite sums, we exclude the zero energy modes, because they only give a contribution of order $1/\sqrt{N}$ which will vanish for large enough system sizes.

\subsection{Strong coupling approximation}
\label{subsec:strong_coupling_approx}
Now to solve the interacting system in the strong coupling regime, we make a Born-Oppenheimer type approximation assuming that the fast dynamics of the meson, consisting of the chargon and the disturbances in the Néel background it creates, decouple from the slower magnon dynamics:
\begin{equation}
|\psi(k)\rangle=|\psi_{mes}(k-\hat{Q}_{a})\rangle\otimes |\tilde{\psi}_{mag}(k-\hat{Q}_{a})\rangle ,
\end{equation}
for fixed LLP-momentum $k$.
The Hilbert space ${\mathcal{H}_{mes}\ni|\psi_{mes}\rangle}$ consists of all possible string configurations (or equivalently the configurations of the Ising variables $\hat{\tau}^{z}_{j}$) and is thus spanned by the truncated basis eigenstates $\lbrace|\phi^{mes}_{n}(k)\rangle\rbrace_{n,k}$. 

The configuration of the magnons introduced in Sec.~\ref{subsec:gen_expansion} can be described by a bosonic Gaussian trial state $|\tilde{\psi}_{mag}(k)\rangle$ in \emph{squeezed space}, characterized entirely by its two-point correlation functions $\langle \hat{a}^{\dagger}_{i}\hat{a}_{j}\rangle$ and $\langle \hat{a}_{i}\hat{a}_{j}\rangle$. Here, \emph{squeezed space} means that the hole motion distorts the magnon configuration along the corresponding string \footnote{Note that by construction every string can be mapped to a unique truncated basis state. The opposite is not true: not every distorted Ising state of $\uparrow,\,\downarrow$ spins can be uniquely associated with a string.} and that the correlations are adapted accordingly; i.e. if the hole hopping leads to the interchange of the spins at sites $j$ and $j+\delta$, the Gaussian magnon state is changed accordingly $\langle \hat{a}_{i}^{(\dagger)}\hat{a}_{j}\rangle\rightarrow\langle \hat{a}_{i}^{(\dagger)}\hat{a}_{j+\delta}\rangle$. This will preserve the Gaussian nature of the magnon state and takes the interplay between the hole motion and the magnons fully into account.

In the next step, we use the slowly-varying magnon approximation and assume that the magnon correlations vary slowly,
\begin{equation}
\langle \hat{a}^{(\dagger)}_{i}\hat{a}_{j+\delta}\rangle \approx \langle \hat{a}^{(\dagger)}_{i}\hat{a}_{j}\rangle\,,
\end{equation}
which is expected to hold at low temperatures, when dressing with long-wavelength, i.e. low-energy magnons dominates.
In this approximation $|\tilde{\psi}_{mag}(k)\rangle \approx |\psi_{mag}(k)\rangle$ the magnon trial state is the Gaussian state in the LLP-frame, with unmodified correlations $\langle \hat{a}^{(\dagger)}_{i}\hat{a}_{j}\rangle$, so that our product ansatz becomes
\begin{equation}\label{eq:product-ansatz}
|\psi(k)\rangle=|\psi_{mes}(k-\hat{Q}_{a})\rangle\otimes |\psi_{mag}(k-\hat{Q}_{a})\rangle .
\end{equation}
This approximation amounts to neglecting the kinetic coupling between the meson and the spin-waves and is only justified because we split each spin degree of freedom into two independent degrees of freedom $\hat{a}_j$ and $\hat{\tau}^z_j$. It is not valid for conventional spin-wave treatments of the $t$-$J$ model within the usual 1/S expansion \cite{Kane1989, Liu1992, Schmitt-Rink1988, Martinez1991, Nielsen2021, Igarashi1992}, where the strong distortion of the AFM background is entirely described by linear spin-wave excitations. In contrast, our formalism captures the correlated hopping processes of the hole and spins by the Ising background $\hat{\tau}_{j}^{z}$ and therefore the coupling of the spin-waves to the hole-motion is weak and involves predominantly long-wavelength magnons.

Now, we can derive an explicit expression of the effective interaction between the meson and magnons.
Making use of the ansatz in Eq.~\eqref{eq:product-ansatz}, we project the interacting part of the Hamiltonian onto the mesonic eigenstates $\lbrace|\phi_{n}^{mes}\rangle\rbrace_{n}$ and get
\begin{equation}
\hat{H}^{int} = \sum_{k,n,n'} \hat{f}_{k-\hat{Q}_{a},n'}^{\dagger}\hat{H}^{int}(k,n,n')\hat{f}_{k-\hat{Q}_{a},n},
\end{equation}
with
\begin{align}
&\hat{H}^{int}(k,n,n')\nonumber\\
=& -J_{z}S\sum_{\langle ij\rangle }\hat{A}_{ij}^{(1)}\nonumber\\
&\times\langle\phi^{mes}_{n'}(k-\hat{Q}_{a})|(\hat{a}^{\dagger}_{i}\hat{a}_{i} +\hat{a}^{\dagger}_{j}\hat{a}_{j})|\phi^{mes}_{n}(k-\hat{Q}_{a})\rangle\nonumber\\
+&\dfrac{J_{\perp}S}{2}\sum_{\langle ij\rangle }\hat{A}_{ij}^{(2)}\nonumber\\
&\times\langle\phi^{mes}_{n'}(k-\hat{Q}_{a})|(\hat{a}^{\dagger}_{i}\hat{a}_{j} +\hat{a}^{\dagger}_{j}\hat{a}_{i})|\phi^{mes}_{n}(k-\hat{Q}_{a})\rangle\nonumber\\
-&\dfrac{J_{\perp}S}{2}\sum_{\langle ij\rangle }\hat{A}_{ij}^{(3)}\nonumber\\
&\times\left(\langle\phi^{mes}_{n'}(k-\hat{Q}_{a})|\hat{a}^{\dagger}_{j}\hat{a}_{i}^{\dagger}|\phi^{mes}_{n}(k-\hat{Q}_{a})\rangle+\textrm{h.c.}\right).
\end{align}
The coefficients are
\begin{align}
\label{coefficients real space A}
\hat{A}_{ij}^{(1)}&=1+\hat{\tau}^{z}_{i}\hat{\tau}^{z}_{j},\\
\label{coefficients real space B}
\hat{A}_{ij}^{(2)}&=1+\hat{\tau}^{z}_{i}\hat{\tau}^{z}_{j}-\delta_{i,0}-\delta_{j,0} ,\\
\label{coefficients real space C}
\hat{A}_{ij}^{(3)}&=1+\hat{\tau}_{i}^{z}\hat{\tau}_{j}^{z}+\delta_{i,0}+\delta_{j,0}+2\hat{F}_{ij}.
\end{align}
Note that $\hat{A}_{ij}^{(1)},\;\hat{A}_{ij}^{(2)}$ and $\hat{A}_{ij}^{(3)}$ are symmetric under exchanging $i\leftrightarrow j$.

After expressing the HP bosons $\hat{a}_{j}$ in terms of magnon operators $\hat{\beta}_{p}$ and using the bosonic commutation relations to normal order the final expression, as well as applying a Fourier transformation, we get
\begin{align}
\label{eff hamiltonian}
&\hat{H}_{int}(k,n,n')=\dfrac{1}{N}\sum_{p\,q}\Big[\nonumber\\
&\langle\phi^{mes}_{n'}(k-\hat{Q}_{a})|\hat{B}_{pq}^{(1)}|\phi^{mes}_{n}(k+p-q-\hat{Q}_{a})\rangle\hat{\beta}_{p}^{\dagger}\hat{\beta}_{q}\nonumber\\
+&\Big(\langle\phi^{mes}_{n'}(k-\hat{Q}_{a})|\hat{B}_{pq}^{(2)}|\phi^{mes}_{n}(k+p+q-\hat{Q}_{a})\rangle\hat{\beta}_{p}^{\dagger}\hat{\beta}_{q}^{\dagger}\nonumber\\
&+\textrm{h.c.}\Big)\nonumber\\
+&\langle\phi^{mes}_{n'}(k-\hat{Q}_{a})|\hat{B}_{pq}^{(3)}|\phi^{mes}_{n}(k-\hat{Q}_{a})\rangle\Big].
\end{align}
Note that the total momentum operator of the Bogoliubov rotated magnons $\hat{Q}_{\beta}=\sum_{k}k\hat{\beta}^{\dagger}_{k}\hat{\beta}_{k}$ is identical to the total momentum operator of the HP magnons ${\hat{Q}_{a}=\sum_{k}k\hat{a}^{\dagger}_{k}\hat{a}_{k}}$.
The explicit expressions for the coefficients $\hat{B}^{(1)},\, \hat{B}^{(2)},\, \hat{B}^{(3)}$ are a bit cumbersome and can be found in appendix \ref{app:coefficients}.
Since the last term above is independent of any magnon operators, we will include it from now on in the free polaron Hamiltonian instead of the $\hat{H}_{int}$. This corresponds to renormalizing the polaron energy
\begin{equation}
E_{n}(k)\rightarrow E_{n}^{*}(k)=E_{n}(k)+\dfrac{1}{N}\sum_{p\,q}\langle\phi^{mes}_{n}(k)|\hat{B}_{pq}^{(3)}|\phi^{mes}_{n}(k)\rangle .
\label{eq:zero-point-renormalization}
\end{equation}

\subsection{Self-consistent Born approximation}
\label{subsec:SCBA}
\begin{figure}[t!]
  \centering\includegraphics[width=1.0\columnwidth]{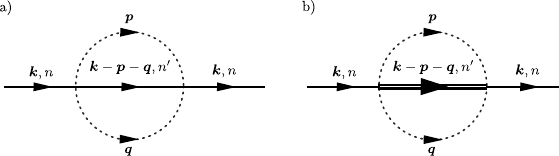}
  \caption{Feynman diagrams for the polaron self-energy in the Born approximation (a) and SCBA (b). The dashed line represents the bare magnon propagator, the simple solid line represents the bare polaron propagator and the double line stands for the full polaron propagator.}
  \label{fig:feynman diagram}
\end{figure}

Now, we analyze how the meson-magnon interaction renormalizes the polaron properties. To this end, we apply the self-consistent Born approximation to compute the hole Green's and spectral functions. This is similar to the procedure used in the conventional 1/S expansion \cite{Kane1989, Liu1992, Schmitt-Rink1988, Martinez1991, Nielsen2021, Igarashi1992}, but starting from the beyond-linear spin-wave Hamiltonian.

We use the zero temperature formalism and define the hole Green's function as usual,
\begin{equation}
G_h(t, k)=-i\langle\psi_0|T\hat{c}_k^{\dagger}(t)\hat{c}_k(0)|\psi_0\rangle,
\end{equation}
where $|\psi_{0}\rangle = |\textrm{Néel}\rangle\otimes|0\rangle_{mag}$ is a product state of the Néel state for the background spins and the vacuum of the magnons, $\hat{\beta}_{p}|0\rangle_{mag}=0$.
Using 
\begin{align}
&\hat{c}_{k}(t)|\psi_0\rangle\nonumber\\
=&\,\sum_{n}\,\langle\psi_0|\hat{f}_{n,k}(t)\hat{c}_{k}(t)|\psi_0\rangle\;\hat{f}_{n,k}^{\dagger}(t)|\psi_0\rangle\nonumber\\
=&\,\sum_{n}\,\langle\psi_0|\hat{f}_{n,k}\hat{c}_{k}|\psi_0\rangle\;\hat{f}_{n,k}^{\dagger}(t)|\psi_0\rangle,
\end{align}
the hole propagator can be expressed in terms of polaron propagators,
\begin{multline}
\label{eq:greens-function-exact}
G_h(t, k)=-i\sum_{n,n'}\langle \psi_0|\hat{f}_{n,k}\hat{c}_{k}|\psi_0\rangle\\
\times\langle \psi_0|\hat{c}_{k}^{\dagger}\hat{f}_{n',k}^{\dagger}|\psi_0\rangle\langle\psi_0|T\hat{f}_{n',k}(t)\hat{f}_{n,k}^\dagger(0)|\psi_0\rangle\,.
\end{multline}
In the following, we use the additional approximation that inter-band processes are only virtually allowed so that $\langle\psi_0|T\hat{f}_{n',k}(t)\hat{f}_{n,k}^\dagger(0)|\psi_0\rangle \sim \delta_{n,n'}$.
The hole propagator then takes the form
\begin{align}
\label{eq:greens-function}
&G_h(t, k)\nonumber\\
&\approx-i\sum_{n}|\langle \psi_0|\hat{f}_{n,k}\hat{c}_{k}|\psi_0\rangle|^2\langle\psi_0|T\hat{f}_{n,k}(t)\hat{f}_{n,k}^\dagger(0)|\psi_0\rangle\nonumber\\
&=\sum_{n}\tilde{Z}_{n}(k)G_{mes}(n, t, k),
\end{align}
where we defined the overlap of the $n$-th meson state with the undisturbed doped Néel state to be $\tilde{Z}_{n}(k) = |\langle \psi_0|\hat{f}_{n,k}\hat{c}_{k}|\psi_0\rangle|^2$ (meson spectral weight).

The dressing of the meson Green's function by the interaction with the spin-waves is captured by the meson self-energy $\Sigma(n, k, \omega)$.
\begin{equation}
G_{mes}(n, k, \omega)^{-1} = G_{mes}^{(0)}(n, k, \omega)^{-1} - \Sigma(n, k, \omega)\,.
\end{equation}
First, we compute the meson self-energy up to the first non-vanishing contributions. This yields
\begin{align}
\label{self energy born}
&\Sigma(n,k,\omega)\nonumber\\
&=\int dp dq \sum_{n'}\;|\langle\phi^{mes}_{n'}(k-p-q)|\hat{B}_{pq}^{(2)}|\phi^{mes}_{n}(k)\rangle|^{2}\nonumber\\
&\quad\times G_{mes}^{(0)}(n',k-p-q ,\omega-\omega^{mag}_{p}-\omega^{mag}_{q})\,,
\end{align}
and is shown in Fig.~\ref{fig:feynman diagram}. All other diagrams up to second order in the coupling $J/t$ vanish at zero temperature.

In the next step, we will make the Born approximation self-consistent by replacing the bare polaron propagator $G_{mes}^{(0)}(n,k,\omega)$ in \eqref{self energy born} by the dressed Green's function $G_{mes}(n,k,\omega)$. We get
\begin{widetext}
\begin{align}
\label{self energy}
\Sigma(n,k,\omega)&=\sum_{n'}\int dp dq	\;|\langle\phi^{mes}_{n'}(k-p-q)|\hat{B}_{pq}^{(2)}|\phi^{mes}_{n}(k)\rangle|^{2}G_{mes}(n',k-p-q,\omega-\omega^{mag}_{p}-\omega^{mag}_{q})\nonumber\\
&=\sum_{n'}\int dp dq	\;\dfrac{|\langle\phi^{mes}_{n'}(k-p-q)|\hat{B}_{pq}^{(2)}|\phi^{mes}_{n}(k)\rangle|^{2}}{\omega-\omega^{mag}_{p}-\omega^{mag}_{q}-E_{n'}^{*}(k-p-q)-\Sigma(n',k-p-q,\omega-\omega^{mag}_{p}-\omega^{mag}_{q})+i\eta}\; .
\end{align}
\end{widetext}
Here $\eta \rightarrow 0^{+}$ is an infinitesimally positive number.

After solving this self-consistent equation iteratively, we can finally compute the renormalized dispersion relation of the polaron $E_{n}^{\textrm{SCBA}}(k)$ by solving
\begin{equation}
\omega - E_{n}^{\textrm{SCBA}}(k) - Re(\Sigma (n, k, \omega )) = 0 \Big|_{\omega =E_{n}^{\textrm{SCBA}}(k)}\, .
\end{equation}
Furthermore, we can get the magnon contribution to the spectral weights of the string states
\begin{equation}
Z_{n}^{-1}(k) = 1-\partial_{\omega}Re(\Sigma(n,k,\omega))\Big|_{\omega=E_{n}^{\textrm{SCBA}}(k)}\, ,
\end{equation}
and compute the spectral function  of the hole
\begin{align}
A_{h}(k,w)&=-\frac{1}{\pi}\textrm{Im}(G_h(k,w))\nonumber\\
&\approx -\frac{1}{\pi}\sum_{n}\tilde{Z}_{n}(k)\textrm{Im}(G_{mes}(n,k,w)).
\end{align}
In the second step, we used the earlier approximation from Eq.~\eqref{eq:greens-function}. We will discuss our results in Sec. \ref{sec:results}.

A comment is in order about the different contributions $\tilde{Z}_{n}(k)$ and $Z_{n}(k)$ to the spectral weight of polaron peaks. The mesonic contribution $\tilde{Z}_{n}(k) = |\langle \psi_0|\hat{f}_{n,k}\hat{c}_{k}|\psi_0\rangle|^2$ is given by the overlaps of the bare hole and the meson. Due to contributions from geometric strings with non-zero length, this factor becomes smaller than one and for instance vanishes for purely rotational meson excitations. The contribution $Z_{n}(k)$ describes the additional decrease of the quasiparticle weight due to the dressing of the meson by spin-waves/magnons.

\section{Results}
\label{sec:results}
Now we present our numerical results obtained using the formalism introduced in Sec. \ref{sec:formalsim}. First, we describe the mesonic part of the magnetic polaron, using the truncated basis (geometric string theory) (see Sec. \ref{subsec:string_picture}). We demonstrate that these mesonic contributions capture all established features of the magnetic polaron dispersion, including the power-law scaling of the ground state energy with $t/J$ as well the location of the dispersion minimum at the nodal point $(\pi/2,\pi/2)$. We then obtain new insights into the excited states of magnetic polarons by (i) discovering an avoided level crossing around the $\Gamma$ point $k=(0,0)$ where we predict the lowest polaron state to be rotationally non-trivial with p-wave character; and (ii) by revealing significant momentum-dependence of the lowest-energy meson state and its string-length distribution.
Second, we analyze the renormalization effects due to the interaction with collective spin-wave excitations/ magnons within SCBA. We demonstrate that this leads to no significant further renormalization of the polaron dispersion beyond the mesonic contribution, whereas the overall polaron energy is more strongly affected. This allows us in a next step to analyze the low-doping pseudogap, originating from the ground state energy difference between the nodal and anti-nodal points. The resulting doping-dependence of the pseudogap we predict is in notable agreement with established phenomenology in the cuprate compounds~\cite{Hashimoto2014}. Using our meson construction, we obtain new insights into the origin of the low-doping pseudogap: We find it to be dominated by Trugman-loop effects when $t \gtrsim 10 \, J$ and by spinon-motion otherwise, whereas dressing with additional magnons leaves the pseudogap essentially unaffected.
Finally, we analyze the spectral function and find that meson-magnon coupling leads to an incoherent peak-dip-hump type structure at low energies above the ground state, and (iii) additional broadening of the dense mesonic spectrum at higher energies.

\subsection{Mesonic contributions:\\
Benchmarking the truncated basis}
\label{subsec:res-trunc-basis-benchmark}
\begin{figure}[t!]
  \centering\includegraphics[width=0.95\linewidth]{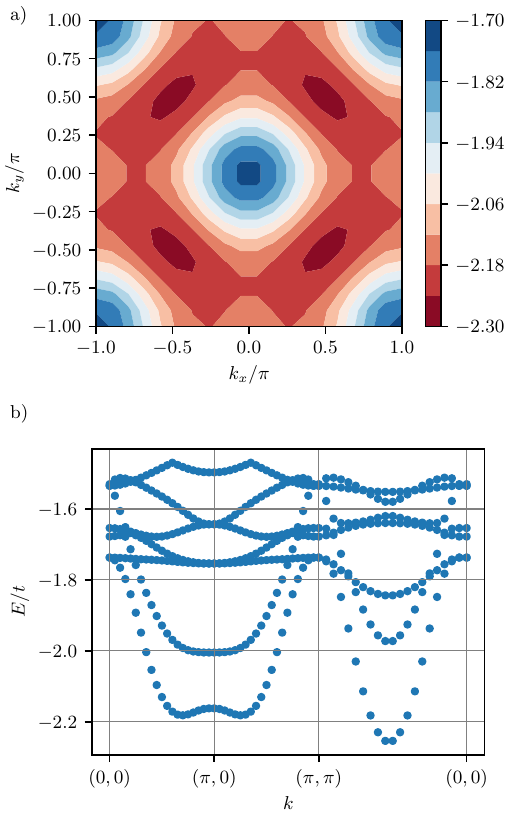}
  \caption{The magnetic polaron in the string picture. The ground state dispersion over the Brillouin zone (a) and the dispersion of the first 7 energy states along a high-symmetry cut of the Brillouin zone (b) are shown. Here we used $l_{max}=8$ and $J/t=0.3$.}
  \label{fig:res-ED-disp}
\end{figure}

\begin{figure}[t!]
  \centering\includegraphics[width=0.95\linewidth]{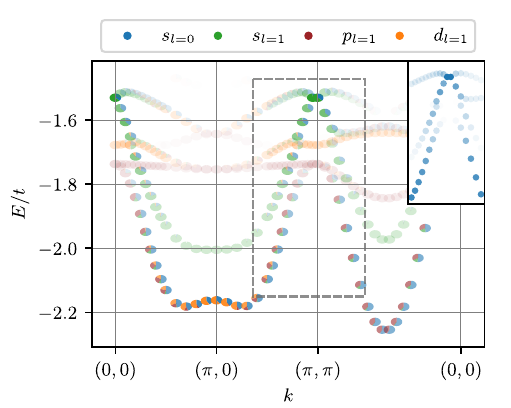}
  \caption{Rotational and vibrational string states. We plot the same energy states as in Fig.~\ref{fig:res-ED-disp} and indicate the contribution of the rotational trial states~\eqref{eq:rot-trial-states} by the filling of the data points. The opacity indicates the sum of the different overlaps and the inset shows the avoided level crossing of the $s_{l=0}$ state in the dashed area near $k=(\pi,\pi)$ by choosing the opacity proportional to the spectral weight $\tilde{Z}_n(k)$.}
  \label{fig:res-ED-disp-rot}
\end{figure}
In Fig.~\ref{fig:res-ED-disp}, we show the dispersion relation of the magnetic polaron when only mesonic states which can be constructed from strings with a maximum length of $l_{max}=8$ are taken into account. We analyze the convergence with the truncation $l_{max}$ in App.~\ref{app:convergence}, where we demonstrate good convergence of relative energy scales for $l_{max} \leq 10$. We find that the truncated basis manages to qualitatively describe the magnetic polaron well, capturing the most important physical processes. In particular, this method captures the dispersion minimum at $k=(\pi/2, \pi/2)$ and the strongly renormalized bandwidth which is only of order $J$ \cite{Kane1989} instead of order $8t$, as would be the case for a free hole.
As we will discuss further below, the truncated meson basis also captures the energy splitting between the nodal $k=(\pi/2, \pi/2)$ and anti-nodal $k=(\pi,0)$ points.

\subsection{Rotational meson ground state at $k=(0, 0)$}\label{subsec:rot-gs}
\begin{figure*}[t!]
  \centering\includegraphics[scale=1.]{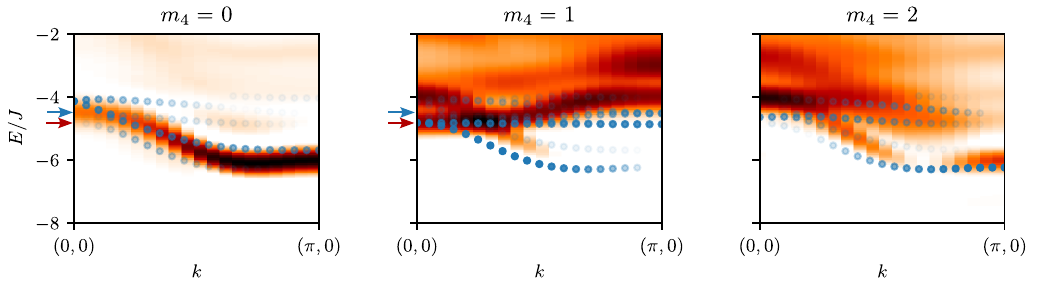}
  \caption{Rotational spectra from DMRG calculations. We show the rotational ARPES spectra and compare our meson picture to DMRG calculations. The numerics were carried out on an elongated $40\times 4$-leg cylinder for $t/J = 3$. The blue (red) arrows indicate the quasiparticle peak at $k=(0,0)$ for $m_4 = 0\;(m_4=1)$ from the DMRG calculations. The blue data points show the string states as obtained in the truncated basis with opacity weighted by the overlap with the corresponding rotational trial states~\eqref{eq:rot-trial-states}.}
  \label{fig:rot-DMRG}
\end{figure*}

Since the meson is an extended object, it can feature rotational and vibrational string excitations. At $C_4$-invariant momenta (C4IM) $k=(0,0)$ and $k=(\pi,\pi)$, we can simultaneously assign linear and angular momentum. This is not possible away from the momenta $k=(0,0),(\pi,\pi)$, because the $C_4$ rotation and translation operators do not commute, and different internal string states can hybridize.

To probe the ro-vibrational nature of the excitations even away from C4IM, we follow Ref.~\cite{Bohrdt2021} and define trial states with well-defined angular momentum \mbox{$m_4\in\lbrace 0,1,2,3\rbrace$} corresponding to s-, p-, d- and f-wave symmetry:
\begin{equation}
\label{eq:rot-trial-states}
|k,m_4, \sigma\rangle = \sum_{j}\dfrac{e^{-ikj}}{\sqrt{N}}\sum_{i:\langle i,j \rangle}e^{im_4 \varphi_{i-j}}\sum_{\sigma'}\hat{c}^{\dagger}_{j, \sigma'}\hat{c}_{i, \sigma'}\hat{c}_{j, \sigma}|\psi_0\rangle\,,
\end{equation}
where $\varphi_{r} = \textrm{arg}(r)$ is the polar angle of $r$. The trial states thus correspond to string states with length $l=1$ and given angular momentum $m_4$.

Fig.~\ref{fig:res-ED-disp-rot} shows again the same energy levels as Fig.~\ref{fig:res-ED-disp}(b) along the high-symmetry cut through the Brillouin zone, but now we indicate the mesonic spectral weight $\tilde{Z}_{n}(k) = |\langle\psi_{0}|\hat{f}_{n,k}\hat{c}_{k}|\psi_{0}\rangle|^2$, corresponding to the overlap with the string length zero state, and the overlap with the rotational trial states \eqref{eq:rot-trial-states}. The overall opacity is determined by the sum of the different overlaps. 

At the C4IM, we find that the lowest mesonic state is degenerate and has angular momentum $m_{4}=1$ or $m_{4}=3$  corresponding to p-wave symmetry and thus zero overlap with the localized hole in the Néel state, i.e. vanishing spectral weight. Away from these momenta, the rotational excitations hybridize and we find signals of avoided level crossing. This is highlighted by the inset of Fig.~\ref{fig:res-ED-disp-rot}, where the energy levels around $k=(\pi,\pi)$ are shown and weighted by their spectral weight.

At the dispersion minimum $k=(\pi/2, \pi/2)$, the ground state shows significant overlap with both s-wave and p-wave trial states. Similarly, the ground state at the anti-nodal point $k=(\pi,0)$ is a hybridization of s and d-wave contributions.

We observe the same features in DMRG calculations of the rotational spectra \cite{Bohrdt2021} shown in Fig.~\ref{fig:rot-DMRG}. These numerical spectra also show the p-wave nature of the ground state at the $\Gamma$-point, i.e. $k=(0,0)$. The lowest energy level then hybridizes with the s-wave state and looses all its p-wave contributions near the anti-nodal point $k=(\pi,0)$. At the same time, the spectrum for $m_4=2$ shows that the lowest energy state at the anti-node has important contributions with d-wave symmetry, which vanish when approaching $k=(0,0)$.
Overall, we find excellent agreement between the truncated basis and DMRG calculations of the rotational excitations.

\begin{figure}[t!]
  \centering\includegraphics[width=0.95\linewidth]{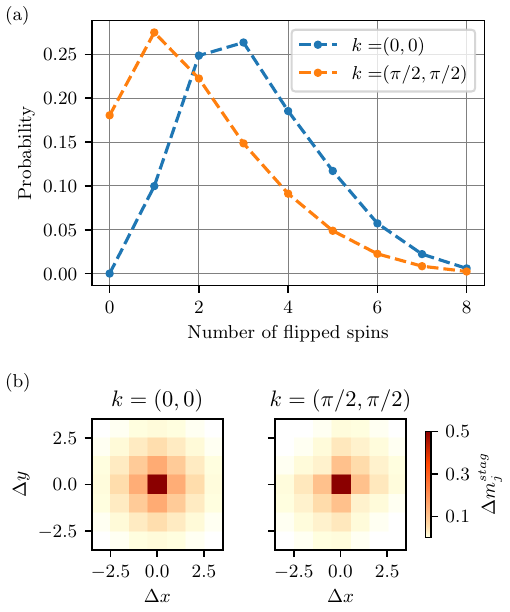}
  \caption{The probability distribution for the number of flipped spins (a) and the staggered magnetization around the hole (b) are shown for ground states at $k=(0,0)$ and $k=(\pi/2, \pi/2)$.}
  \label{fig:magnetization}
\end{figure}

The different character of the polaron ground state at $k=(0,0)$ and $(\pi/2, \pi/2)$ is also reflected in the experimentally accessible shape of the polaron cloud around the hole. In Fig.~\ref{fig:magnetization}(a) the probability distribution for the string length calculated in the truncated basis is shown. We observe that for $k=(0,0)$ the undisturbed string-length $l=0$ state does not contribute to the ground state.
Furthermore, the probability distribution shows that while at the dispersion minimum $(\pi/2, \pi/2)$ states with only one flipped spin contribute the most, the rotational string state at $k=(0,0)$ has larger contributions from longer string states. Both observations directly reflect the node forming around $l=0$ in the rotational ground state. This can also be seen from the staggered magnetization $\Delta m^{stag}_{j} = S-(-1)^{j}\langle S_{j}^{z}\rangle$ shown in Fig.~\ref{fig:magnetization}(b), since at $k=(0,0)$, the polarization of the magnetic background is stronger and further extended compared to $k=(\pi/2, \pi/2)$. In addition, this figure shows that the truncated basis methods allows to access the real-space structure of the polaron and manages to capture the anisotropy of the polarization cloud at $k=(\pi/2, \pi/2)$, agreeing qualitatively with predictions by linear spin-wave calculations in~\cite{Nielsen2021}.

We want to note that the physics near $k=(0,0)$ is very hard to extract because of different processes competing at very similar energy scales. The bandwidth of the free spin-waves $W_{mag}^{(0)} = 2J$ is almost identical to the bandwidth of the magnetic polaron and if we choose a different truncation scheme allowing for disconnected geometric strings (see Sec.~\ref{subsec:string_picture} and App.~\ref{app:connected-strings}), the zero angular momentum state gets shifted to lower energies and we do not observe any avoided level crossing. Nevertheless, we believe this level crossing to be physical since it agrees with DMRG data and could possibly explain the jump of the bandwidth obtained with QMC methods by Brunner et al.~\cite{Brunner2000}, as discussed below around Fig.~\ref{fig:bandwidth}.

\subsection{Magnon dressing: Benchmarking beyond-linear spin-wave SCBA}
\label{subsec:res-magnon}
Now we include the collective spin-wave excitations as introduced in \ref{subsec:gen_expansion} and present the effect of the string's dressing by these magnons. The following results have all been obtained by approximating the momentum integrals as sums on a reciprocal lattice of $12\times 12$ sites and by including the coupling of the first $7$ internal mesonic states. A discussion of the convergence with the number of internal mesonic states can be found in App.~\ref{app:conv-nmax}.

\begin{figure}[t!]
  \centering\includegraphics[width=1\linewidth]{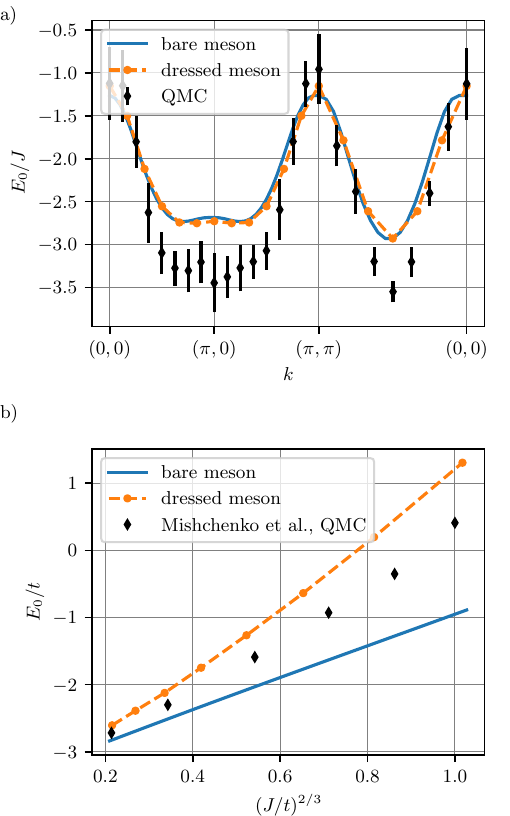}
  \caption{The dressing of the meson by the spin-waves. (a) We show the renormalized dispersion relation for $J/t=0.4$ and compare to the undressed case and QMC data \cite{Brunner2000}. An overall energy shift has been applied to the bare meson dispersion.
(b) The ground state energy as a function of $(J/t)^{2/3}$ is shown and compared to the undressed case and QMC data \cite{Mishchenko2001}. The ground state energies are shown in units of $t$ with respect to the undoped Heisenberg antiferromagnet.}
  \label{fig:scba_disp}
\end{figure}

First of all, we compare the dispersion relation of the bare and dressed meson in Fig.~\ref{fig:scba_disp}(a) at $J = 0.4 \, t$. While the overall energy experiences a shift, we focus on the shape of the dispersion and measure energy relative to the ground state at $k=(\pi/2, \pi/2)$. We find that the shape of the dispersion remains almost unchanged for $J<t$. Additionally, we compare to QMC simulations by Brunner \emph{et al.}~\cite{Brunner2000} and find that while the qualitative features of the dispersion and even the energies at the upper part of the bandwidth agree quite well, our dressed meson theory predicts slightly higher ground state energies and thus a smaller bandwidth.

In the panel (b) of Fig.~\ref{fig:scba_disp}, we show the ground state energy as a function of $(J/t)^{2/3}$ and compare to QMC data from Mishchenko \emph{et al.} \cite{Mishchenko2001}. We see that the meson in the geometric string picture has energies lower than predicted by the QMC simulations. This is not a physical feature but instead due to the fact that we compute all energies with respect to the ground state of the undoped AFM. In the string picture however, the undoped antiferromagnet has perfect Néel order and displaying the energies with respect to the Néel state leads to too low values because it has higher energy than the true ground state. Including the spin-waves on the other hand takes the spin fluctuations of the ground state into account and we obtain energies which are higher than and closer to the QMC simulations. Nevertheless, all three methods show that the ground state energy scales as $J^{2/3}\,t^{1/3}$. This is due to the linear string tension and derivations can be found for instance in \cite{Grusdt2018, Bulaeski1968}.

\begin{figure}[t]
  \centering\includegraphics[width=1\linewidth]{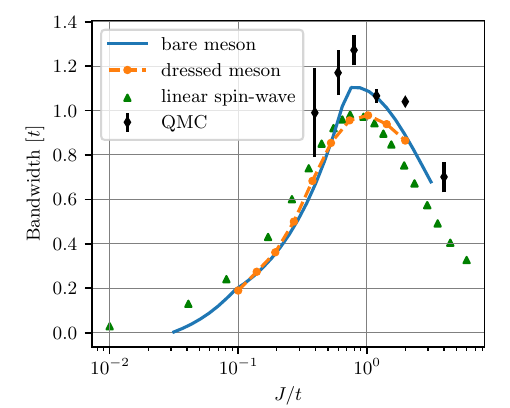}
  \caption{The bandwidth as a function of $J/t$ compared with results from QMC methods \cite{Brunner2000} and linear spin-wave theory \cite{Martinez1991}.}
  \label{fig:bandwidth}
\end{figure}

Next, we show the bandwidth  in Fig.~\ref{fig:bandwidth} as functions of the coupling $J/t$ and compare again to the bare results as well as state of the art numerical results in the form of QMC simulations by Brunner \emph{et al.} \cite{Brunner2000} and the conventional $1/S$ expansion as computed for instance by Martinez \emph{et al.} \cite{Martinez1991}. This quantity is independent of overall energy shifts, therefore allowing for a better comparison of the different methods and we obtain good overall agreements.


First of all, we remark, that the bandwidth features kinks at both $J/t \approx 0.1$ and $J/t\approx 0.7$. These kinks are due to level crossings of different rotational meson states which scale differently with the spin coupling parameter $J$. More precisely, the lowest energy string ground state at $k=(0,0)$ changes its nature in the interval between the kinks $0.1 \lesssim J/t \lesssim 0.7$, and corresponds to a rotational configuration of the meson with $C_4$-eigenvalue $m_4 = 1, 3$ (p-, f-wave) and vanishing spectral weight $\tilde{Z}_{0}(k=(0,0))=0$ (see Sec.~\ref{subsec:rot-gs}), while outside of this interval, the ground state for zero momentum will have finite spectral weight and $C_4$-eigenvalue $m_4=0$ (s-wave). More details about the scaling of the mesonic states with different rotational symmetries can be found in App.~\ref{app:scaling}.

These level crossings could also potentially be related to the kink possibly observed in the Monte Carlo simulations around $J \approx t$.
In Ref.~\cite{Brunner2000}, they first use a QMC algorithm to compute the imaginary-time Green's function $G(k, \tau)$ and then extract the ground state energy from this function. However, \eqref{eq:greens-function} shows that the Green's function cannot capture the rotational string states because of the vanishing spectral weight and we therefore suppose that their algorithm doesn't see the lowest energy state at $k=(0,0)$ or $k=(\pi,\pi)$ but sees the first excitation with non-vanishing spectral weight. Hence, we assume that the real bandwidth lies below the first three QMC data points.

In addition to QMC, we compare our results to the conventional $1/S$ expansion as has been used in \cite{Kane1989, Liu1992, Schmitt-Rink1988, Martinez1991, Nielsen2021, Igarashi1992}, which gives very similar results.

\begin{figure}[t]
  \centering\includegraphics[scale=0.95]{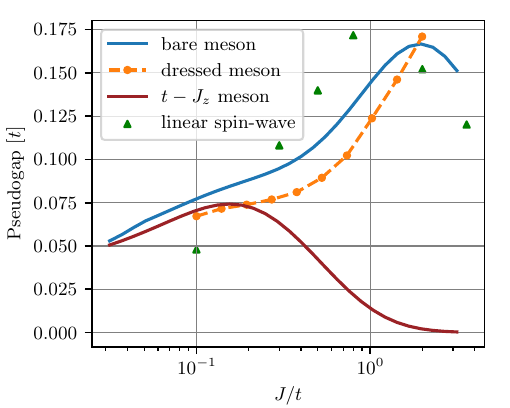}
  \caption{The zero-doping pseudogap, defined by comparing nodal and anti-nodal polaron energy, as a function of $J/t$. We compare to results from linear spin-wave theory \cite{Martinez1991} and to the $t-J_z$ model with Ising spins, where the pseudogap is dominated by Trugman loops.}
  \label{fig:pseudogap}
\end{figure}

\subsection{Low-doping pseudogap}
\label{subsec:res-pseudogap}
So far, we have presented the magnetic polaron as elementary excitation at zero doping and experiments \cite{Kurokawa2023} show that this perspective remains valid to describe cuprates at doping up to $5\%$ and possibly beyond. At finite doping, the magnetic polarons then form a Fermi sea around the nodal point $k=(\pi/2, \pi/2)$ and the disputed pseudogap opening at the anti-nodal point $k=(\pi, 0)$ is hence interpreted as the usual gap present in Fermi liquid theory for excitations away from the Fermi surface,
\begin{equation}
\label{eq:pseudogap}
\Delta_{PG} = E(\pi, 0) - E_F\,.
\end{equation}

\begin{figure}[t]
  \centering\includegraphics[scale=0.95]{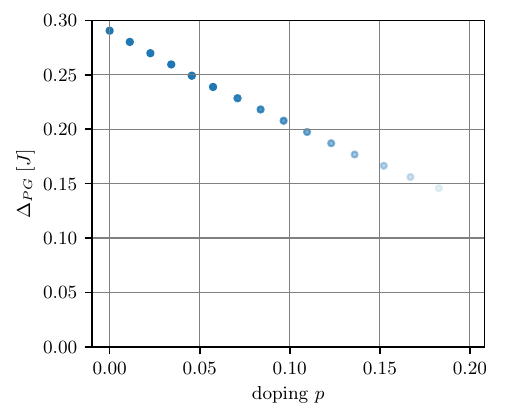}
  \caption{The low-doping pseudogap, defined as difference of the anti-nodal polaron and Fermi energies, as a function of doping $p$. We used $t/J = 3$ and bare meson energies. The decreasing opacity indicates the breakdown of the meson model for higher doping.}
  \label{fig:pseudogap-doping}
\end{figure}

\subsubsection{Zero-doping pseudogap}

\begin{figure*}[t!]
  \centering\includegraphics[scale=1]{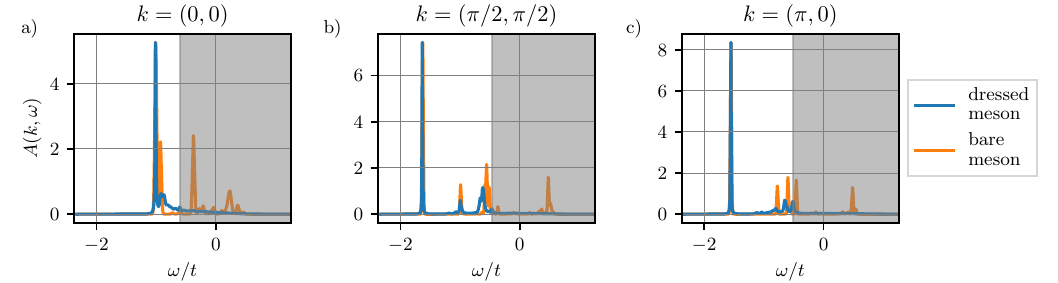}
  \caption{The spectral function of the magnetic polaron at momenta $k = (0,0)$ (a), $k = (\pi/2,\pi/2)$ (b) and $k = (\pi,0)$ (c) for $J = 0.3\,t$. To compute the spectrum of the dressed meson, the first 36 string states were used and higher excitations in the grey area were not resolved. In the case of the bare meson, the first $300$ excited states were used and its spectrum has been artificially broadened and shifted to match the height and position of the lowest-energy quasiparticle peak of the dressed meson.}
  \label{fig:spectral-function}
\end{figure*}

First, we show the pseudogap at zero doping in Fig.~\ref{fig:pseudogap} for the same models as above and additionally compare it to the meson in the $t-J_{z}$ model: $\Delta_{PG} = E(\pi, 0) - E(\pi/2, \pi/2)\,.$
In the latter model consisting of Ising spins, the energy gap between nodal $(\pi/2, \pi/2)$ and anti-nodal $(\pi, 0)$ points is primarily due to the possibility of the polaron to move through the system without disturbing the magnetic background by going through Trugman loops~\cite{Trugman1988}. The smallest of the Trugman loops consist of going around a unit plaquette one and a half times and allow the hole to diagonally hop to next-nearest neighbours (NNN) which contributes to the meson dispersion and lifts the ground state degeneracy. In Ref.~\cite{Grusdt2018}, where the magnetic polarons are described as bound spinon-chargon pairs, a tight-binding model was proposed, where the Trugman loops for the chargon lead to diagonal NNN hopping of the spinon. The resulting tight-binding dispersion of the polaron features a gap between the nodal and anti-nodal points. Even when generalizing to the $t-J$ model and including the transverse spin fluctuations $J_{\perp}$ in a similar perturbative tight-binding prescription for the spinon \cite{Grusdt2019}, the $J_{\perp}$ contribution to the dispersion is degenerate along the boundary of the magnetic Brillouin zone and the pseudogap is solely determined by Trugman loop effects.

By comparing the pseudogap in Fig.~\ref{fig:pseudogap} for Ising and Heisenberg spins, we find that at low spin coupling $J/t\lesssim 0.2$, the pseudogap does not significantly change when adding transverse spin fluctuations $\sim J_{\perp}$. We therefore conclude that the pseudogap is mostly a consequence of the Trugman loops in the strong coupling limit $J\ll t$.

In addition, the importance of the Trugman loops becomes apparent in Fig.~\ref{fig:conv-disp} (in App.~\ref{app:convergence}) showing that strings of length $3$ or longer are necessary to capture the pseudogap. At the same time, $3$ is the minimal string-length able to capture the effect of Trugman loops. Since the smallest of these loops have length $6$, there exist some pairs of strings with length $3$ which are identical up to a Trugman loop, while string configurations of length $2$ can only be related to configurations of length $4$ or higher via Trugman loops, see also \cite{Grusdt2018}.
This highlights again the importance of loop effects for the pseudogap.

For stronger spin coupling $J$, the effect of the Trugman loops becomes weaker, since the strings (meson size) becomes smaller but in the $t-J$~model, the pseudogap still increases. This shows that perturbative treatments of the spin flip-flop processes are insufficient and the dependence of the meson wave function on the transverse spin fluctuations $J_\perp$ becomes important. We find that this effect is already captured by the bare meson and the coupling to the generalized spin-waves does not lead to any important quantitative changes.

\begin{figure*}[t!]
  \centering\includegraphics[scale=1]{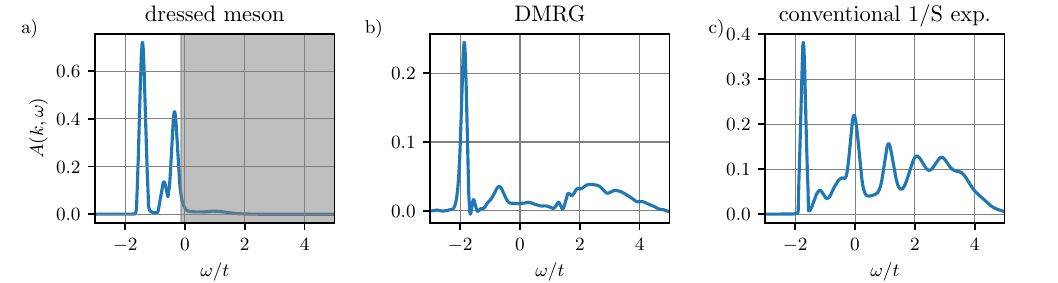}
  \caption{The spectral function of the magnetic polaron at momentum $k=(\pi/2, \pi/2)$ for $J/t = 1/3$ in the dressed meson picture (a), computed with DMRG methods (b) and computed within the conventional $1/S$ expansion (c). In panel (a) and (c), the peaks were broadened artificially to better compare to the DMRG results. }
  \label{fig:spectral-function-node}
\end{figure*}

\subsubsection{Finite-doping pseudogap}
As explained above, from the magnetic polaron perspective, we interpret the pseudogap as excitation gap of the magnetic polaron at the anti-nodal point $k=(\pi,0)$, given by Eq.~\eqref{eq:pseudogap}. In the low doping regime of $p\leq 5\,\%$, where the magnetic polaron description should remain valid \cite{Kurokawa2023}, we can then compute the Fermi energy as a function of doping $p$ and extract the pseudogap as shown in Fig.~\ref{fig:pseudogap-doping}. We observe that the pseudogap decreases linearly with doping, and the order of magnetitude as well as the qualitative behaviour are in good agreement with ARPES measurements~\cite{Hashimoto2014}. Note that this estimate for the pseudogap does not include interactions between the individual magnetic polarons and thus is only valid for dilute polaron gases and breaks down for higher doping, when the magnetic polarons start to overlap and antiferromagnetism diminishes.

\subsection{Spectral function}
\label{subsec:res-spectral-function}
Finally, Fig.~\ref{fig:spectral-function} shows the low-energy part of the spectral function of the magnetic polaron. We can clearly identify the sharp quasiparticle peaks with finite weight as well as the first few excitations which correspond to ro-vibrational excitations of the geometric string (bare meson). We see that the shape of the spectral function is for the most part determined by the mesonic excitations. The dressing by generalized spin-waves gives only comparatively small additional features and reduces the spectral weight of the mesonic excitations.

Even though the spectral function is mostly determined by the bare meson in the geometric string picture, it differs significantly from the spectral function computed in the linear string theory \cite{Grusdt2018}.
In contrast to the ladder-like spectrum in the linear string theory, the bare meson picture, which includes effect of non-linear strings and loops as well as spin flip-flop processes non-perturbatively, does not lead to these equally spaced resonances and is in better agreement with more exact studies~\cite{Bohrdt2020, Brunner2000, Mishchenko2001}.

In addition, we compare the spectral function above to results from the conventional $1/S$ expansion and DMRG in Fig.~\ref{fig:spectral-function-node}. It shows that while the conventional $1/S$ expansion can capture the quasiparticle peak and scattering continuum at high energies quite well, it overestimates the weight of the excitation resonances and features some excitations peaks which are completely absent in the DMRG calculations. Furthermore, the $1/S$ expansion overestimates the excitation energy of the first vibrational peak. In contrast, our meson picture gives an estimate for the excitation energy of the first resonance which is in much better agreement to the full numerical spectrum. However, the method still overestimates the weight of said resonance. We suspect that this is due to the approximation that couplings between internal mesonic states are only virtually allowed \eqref{eq:greens-function}, which does not account for the decay of the high energy states.
After the vibrational excitation peak, we see the onset of a magnon scattering continuum, but further high-energy excitations and scattering processes were not resolved due to the limited number of mesonic string states included in the computations.

\begin{figure}[t]
  \centering\includegraphics[scale=0.95]{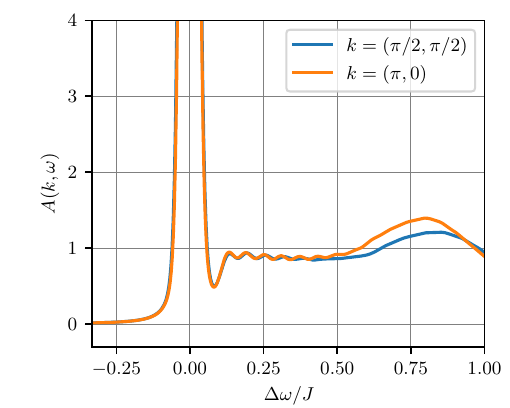}
  \caption{The low-energy part of the spectral function at the nodal and anti-nodal points relative to the quasiparticle energy for $J = 0.3\,t$.}
  \label{fig:low-energy}
\end{figure}

However, note that the limited resolution of the considered momentum grids of $12 \times 12$ sites leads to finite-size effects and the lowest-energy considered spin-waves (apart from the zero-energy modes) have already sizeable energies.
To see the effect of the dressing by low-energy spin-waves, we use the simple low-energy effective model, where we sum only over momenta in a smaller area around the nodal or anti-nodal point and neglect all processes including magnons with momenta outside of this area, since they have higher energy anyway. Here we choose a reciprocal grid of $19 \times 19$ sites with distance $\Delta k = 0.04$, so that the minimal spin-wave energy considered is $\omega_{min}\approx 0.057\,J$ and the maximal included spin-wave energy is $\omega_{max}\approx 0.7\,J$. In this area, we can sum over a much finer grid and obtain the spectral functions shown in Fig.~\ref{fig:low-energy}. We again observe a very sharp quasiparticle peak followed by a tail describing the spin-wave continuum.  Notably, the latter features a peak-dip-hump type structure on an energy scale set by $J$.
We thus conclude that the low-energy spin-waves add an incoherent continuum but do not change the structure of the spectral function in Fig.~\ref{fig:spectral-function}.

This access to the excitation spectrum is one of the main advantages of this model since previous methods have not been able to qualitatively achieve this. Due to the ill-defined problem of analytic continuation, QMC methods cannot resolve the individual string excitations but only see a very "coarse-grained" spectrum. The conventional $1/S$ expansion on the other hand also gives access to the spectrum, but due to the higher level of approximation, the resonances above the quasiparticle peak appear at energies which are too high, when compared to the coarse-grained QMC spectra. Coarse-graining and comparing our spectrum to the QMC spectra leads however to a good agreement apart from an overall energy shift. This thus opens new possibilities to analyze transport properties and non-equilibrium scenarios, where the excitations spectrum of the polaron becomes important.

\section{Summary and Outlook}
We have investigated magnetic polarons in doped antiferromagnets and developed a new formalism valid in the strong coupling regime $t > J$. We made use of the separation of time scales in this regime to decouple the spin dynamics induced by the fast hole motion from the slower transverse spin fluctuations. The fast and strongly correlated hole and spin hopping processes were described using geometric strings and treated with exact diagonalization for a precise description of the dominant contributions. We have shown that the truncated Hilbert space, consisting of a subset of these string states, qualitatively captures the magnetic polaron's properties up to an overall energy shift. Furthermore, we identified an avoided crossing of different internal string states near the center of the Brillouin zone that is hard to capture in many other methods such as QMC \cite{Brunner2000,Mishchenko2001} or conventional linear spin-wave theory \cite{Kane1989, Liu1992, Schmitt-Rink1988, Martinez1991, Nielsen2021, Igarashi1992}.
The remaining processes were treated in linear spin-wave theory using a generalized $1/S$ expansion and we found that those couple only weakly to the pre-formed polaron (or meson), consisting of the hole and its geometric string. This allowed us to derive an effective Hamiltonian of the magnetic polaron with a density-density interaction between the spin-waves and the meson. We then made use of the self-consistent Born approximation to solve the interacting model and showed that the dressing by spin-waves only leads to small quantitative changes. The main advantage of the theory developed here could then be seen from the spectrum of the magnetic polaron. While many previous methods can extract the ground state energy and properties quite well, they have troubles accessing excited states and thus cannot qualitatively resolve the spectral function in contrast to the formalism developed here.

The effective meson-magnon polaron Hamiltonian, which is the main result of this work, connects the problem of doped Mott insulators to the well-known weakly coupled Bose polaron model. This makes the application of specialized methods developed in this field possible. Therefore, our theoretical formalism paves the way for several future applications and allows to study non-equilibrium dynamics of mobile holes and transport properties of doped antiferromagnets as well as finite temperature properties.
Apart from the application for the single polaron, this formalism is quite flexible and allows for the study of different lattice geometries including frustrated lattices and could be modified for the study of bound hole pairs which will be the subject future works.

\section{Acknowledgments}
We thank Lukas Homeier and Jan-Philipp Christ for fruitful discussions. -
This project has received funding from the European Research Council (ERC) under the European Union’s Horizon 2020 research and innovation programm (Grant Agreement no 948141) — ERC Starting Grant SimUcQuam, and from the Deutsche Forschungsgemeinschaft (DFG, German Research Foundation) under Germany's Excellence Strategy -- EXC-2111 -- 390814868.

%

\appendix

\section{Connected vs. unconnected strings}
\label{app:connected-strings}
In the construction of the truncated basis of string states in Sec. \ref{subsec:string_picture}, we chose to consider only states where the hole and flipped spins form a single connected component and describe all flipped spins which cannot be combined into this connected string with the help of the generalized magnons \eqref{eq:HP_operators}.
In this way, we do not attribute all of the spin dynamics around the hole to the hole motion, which reduces the double-counting of $J_{\perp}$-processes when introducing spin-waves.
We further justify the elimination of unconnected strings, because it results in a better agreement with DMRG data obtained by Bohrdt et al. \cite{Bohrdt2021, Bohrdt2021arxiv} when comparing energies at momentum $k=0$.

Note that this restriction on the possible string states is also necessary for energies to converge with the maximal string length $l_{max}$. If we included all string states, more and more transverse spin fluctuations would be captured with the truncated basis. But it is known that for undoped AFMs, these transverse spin fluctuations lower the energy of the ground state by an amount proportional to the volume of the system \citep{Auerbach1998} and therefore all energies would keep on decreasing for growing $l_{max}$.

\section{Convergence of the truncated string basis}
\label{app:convergence}
\begin{figure}[t!]
  \centering\includegraphics[scale=1]{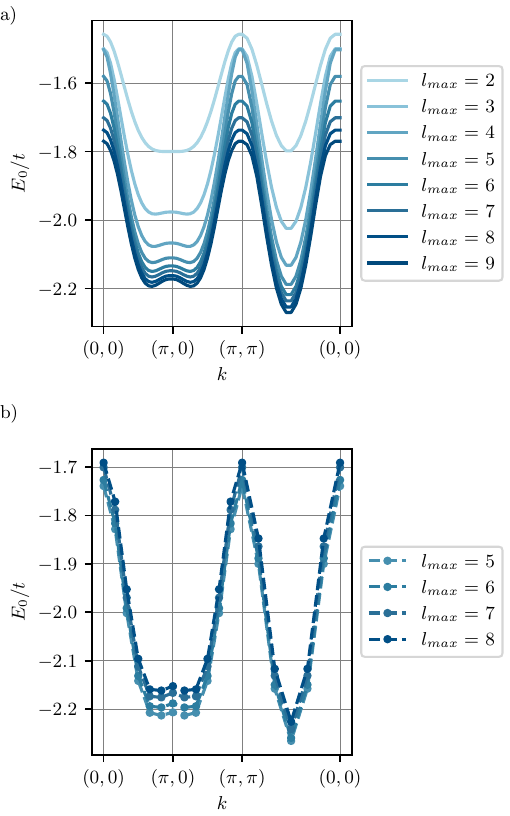}
  \caption{The dispersion of the magnetic polaron for different truncations $l_{max}$ in the geometric string picture (a) and with coupling to spin-waves (b) for $J/t = 0.3$.}
  \label{fig:conv-disp}
\end{figure}
Here we analyze the dependence of the truncation of the geometric string Hilbert space on the magnetic polaron's properties. Fig.~\ref{fig:conv-disp} shows the dispersion of the magnetic polaron for different truncations $l_{max}$ first without coupling to the spin-waves and then including them in the SCBA. First of all, we want to draw attention to the case where $l_{max}=2$. Here we can see that the energy at the nodal point $k=(\pi/2,\pi/2)$ and the anti-node $k=(\pi,0)$ are identical. In fact, the ground state energy is degenerate along the entire edge of the magnetic Brillouin zone for $l_{max}=2$. As mentioned in the main part of the paper, this degeneracy gets lifted due to the Trugman loops \cite{Trugman1988}. But since the shortest of these loops has length $6$, we need $l_{max}\geq 3$ to form different strings which are identical up to a Trugman loop and lift the degeneracy between the node and anti-node.

Second, we find that for larger Hilbert spaces, the overall energies decrease and no clear convergence is obtained. Nevertheless, relative energies and hence the overall shape and bandwidth of the dispersion do converge with growing $l_{max}$.
When we include the coupling between the meson and spin-waves, the overall shape of the dispersion cannot be distinguished for the different truncations shown in Fig.~\ref{fig:conv-disp}, but the convergence of the ground state energy is slower and the energy increases with growing $l_{max}$. This might seem counterintuitive at first, but the collective transverse spin fluctuations which lower the overall energy in the undoped AFM get suppressed around the hole due to distortion of the magnetic background, thus raising the energy of the magnetic polaron.

\begin{figure}[t!]
  \centering\includegraphics[width=1\linewidth]{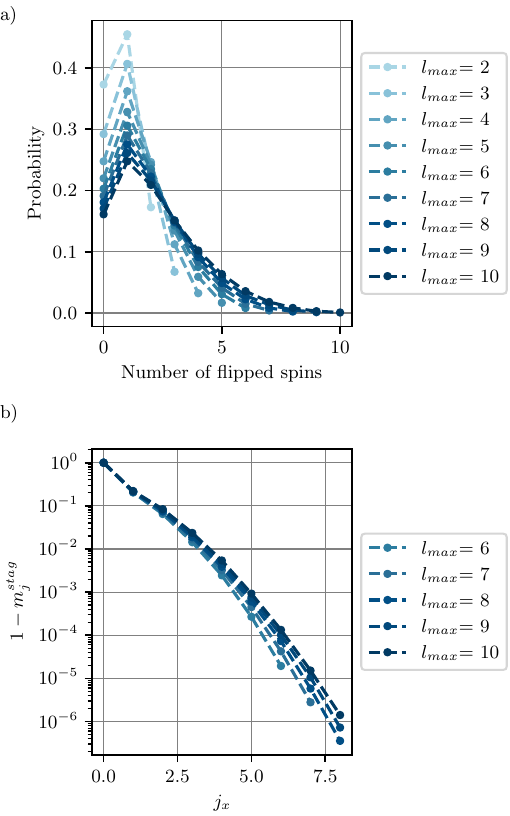}
  \caption{Convergence of the local magnetization (a) and distribution of the number of flipped spins (b) with the size of the truncated basis for $k=(\pi/2, \pi/2)$.}
  \label{fig:conv-mag}
\end{figure}

Furthermore, we consider the effect of the truncation on the spin background of the mesonic ground state and plot the distortion of the local staggered magnetization around the hole as well as the distribution of the number of flipped spins in Fig.~\ref{fig:conv-mag}. On the one hand, the probability distribution for the string lengths is again not entirely converged for the truncations considered here. On the other hand, the local magnetization around the hole only changes weakly for $l_{max}\geq 6$.

Overall, the ground state energy is challenging to predict quantitatively, but all relative energy differences converge quite well. Therefore, energy differences such as the bandwidth and excitation energies, as well as local observables are expected to give qualitatively accurate results.

\section{Coefficients in the effective Hamiltonian}
\label{app:coefficients}
Here we give the explicit expressions of the coefficients in Eq.~\eqref{eff hamiltonian}. They are obtained from the expressions for the coefficients in Eqs \eqref{coefficients real space A}-\eqref{coefficients real space C} after applying a Fourier and a Bogoliubov transformation:
\begin{align}
\hat{B}_{p,q}^{(1)}=&(u_{p}u_{q}+v_{p}v_{q})\nonumber\\
&\times\Big[-J_{z}S\sum_{\langle i,j\rangle}\hat{A}_{ij}^{(1)}(e^{i(q-p)i}+e^{i(q-p)j})\nonumber\\
&\quad+\dfrac{J_{\perp}S}{2}\sum_{\langle i,j\rangle}\hat{A}_{ij}^{(2)}(e^{i(qj-pi)}+e^{i(qi-pj)})\Big]\nonumber\\
&+(u_{p}v_{q}+u_{q}v_{p})J_{\perp}S\sum_{\langle i,j\rangle}\hat{A}_{ij}^{(3)}e^{i(qj-pi)},
\end{align}
\begin{align}
\hat{B}_{p,q}^{(2)}=&-\dfrac{1}{2}(u_{p}v_{q}+u_{q}v_{q})\nonumber\\
&\times\Big[-J_{z}S\sum_{\langle i,j\rangle}\hat{A}_{ij}^{(1)}(e^{-i(q+p)i}+e^{-i(q+p)j})\nonumber\\
&\quad+\dfrac{J_{\perp}S}{2}\sum_{\langle i,j\rangle}\hat{A}_{ij}^{(2)}(e^{-i(qj+pi)}+e^{-i(qi+pj)})\Big]\nonumber\\
&-(u_{p}u_{q}+v_{p}v_{q})\dfrac{J_{\perp}S}{2}\sum_{\langle i,j\rangle}\hat{A}_{ij}^{(3)}e^{-i(pi+qj)},
\end{align}
\begin{align}
\hat{B}_{p,q}^{(3)}=&\delta_{p,q}v_{p}^{2}\nonumber\\
&\times\Big[-2J_{z}S\sum_{\langle i,j\rangle}\hat{A}_{ij}^{(1)}\nonumber\\
&\quad+\dfrac{J_{\perp}S}{2}\sum_{\langle i,j\rangle}\hat{A}_{ij}^{(2)}(e^{ip(j-i)}+e^{ip(i-j)})\Big]\nonumber\\
&+\delta_{p,-p}\,u_{p}v_{p}\,J_{\perp}S\sum_{\langle i,j\rangle}\hat{A}_{ij}^{(3)}e^{ip(i-j)}.
\end{align}

With the expressions above, we can write the coefficients in Eq.~\eqref{eq:pol-Hamiltonian} as:
\begin{align}
B^{(1)} &= B^{(1)}(k,p,q,n,n')\nonumber\\
&=\langle\phi^{mes}_{n'}(k)|\hat{B}_{pq}^{(1)}|\phi^{mes}_{n}(k+p-q)\rangle\,\\
B^{(2)} &= B^{(2)}(k,p,q,n,n')\nonumber\\
&=\langle\phi^{mes}_{n'}(k)|\hat{B}_{pq}^{(2)}|\phi^{mes}_{n}(k+p+q)\rangle\,.
\end{align}

\section{Scaling of energy levels}
\label{app:scaling}

\begin{figure}[t!]
	\includegraphics[scale=1.0]{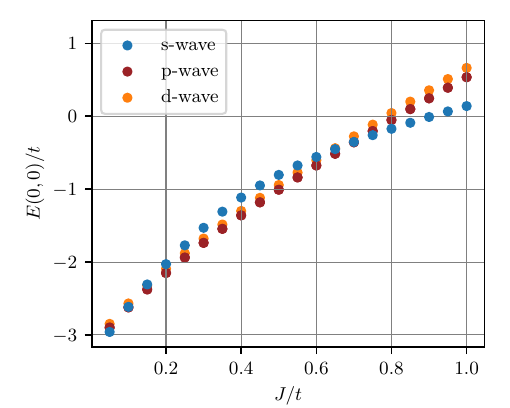}
	\caption{The scaling of the energy of rotational string states. We plot the energy of the first mesonic eigenstates at momentum $k=(0,0)$ with discrete $C_4$-rotational eigenvalues $m_4=0$ (s-wave), $m_4=1$ (p-wave) and $m_4=2$ (d-wave) as a function of $J/t$.}
	\label{fig:energy-scaling}
\end{figure}

In Fig.~\ref{fig:energy-scaling}, we show the different scaling of the mesonic string states at zero momentum as a function of the spin coupling $J/t$ and find that the energy level with discrete $C_4$-rotational eigenvalues $m_4=0$ scales differently with the coupling strength $J/t$ than energy levels with $m_4 \neq 0$.
This leads to two level crossings, where the rotational symmetry of the lowest energy state changes. Between $J/t \approx 0.1$ and $J/t \approx 0.7$, the lowest energy state at $k=(0,0)$ shows p-wave symmetry while it features s-wave symmetry in rest of the scanned parameter region.
These level crossings lead to the kinks observed in Fig.~\ref{fig:bandwidth} and described in Sec.~\ref{subsec:res-magnon}.
\newpage
\begin{figure}[t]
	\includegraphics[scale=1.0]{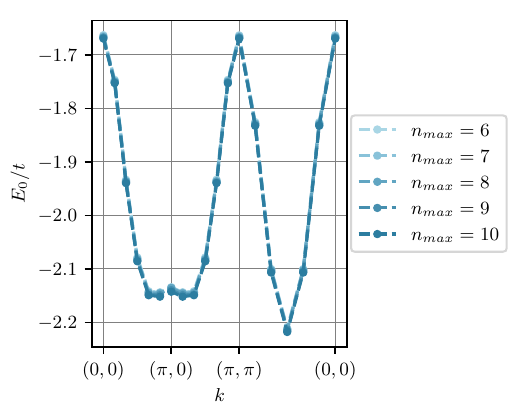}
	\caption{The dispersion of the magnetic polaron for different truncations $n_{max}$ and $J/t = 0.3$. Here, $n_{max}$ denotes the number of internal mesonic energy levels coupling to the lowest energy level.}
	\label{fig:conv-nmax}
\end{figure}

\section{Convergence with the number of excited mesonic states}
\label{app:conv-nmax}

In the computation of the polaron self-energy~\eqref{self energy}, we allow for virtual internal excitations of the meson. Throughout the paper, when we look at ground state properties, we include the coupling between the first $7$ mesonic states an drop the interactions between higher mesonic excitations. Fig.~\ref{fig:conv-nmax} shows the dispersion of the magnetic polaron for different values of the number of included mesonic configurations $n_{max}$. While it is important to take the first few string excitations into account, because of the level crossing of the rotational mesonic configurations, we cannot really distinguish between the different values for $n_{max}$ in Fig.~\ref{fig:conv-nmax} and conclude that $n_{max}=7$ is sufficient to reach convergence.
\end{document}